\documentclass[journal]{IEEEtran}
\IEEEoverridecommandlockouts
\usepackage{cite}
\usepackage{amsmath,amssymb,amsfonts}
\usepackage{graphicx}
\usepackage{multirow}
\usepackage{multicol}
\usepackage{textcomp}
\usepackage{xcolor}
\usepackage{hyperref}
\usepackage{makecell}
\usepackage{soul}
\def\BibTeX{{\rm B\kern-.05em{\sc i\kern-.025em b}\kern-.08em
    T\kern-.1667em\lower.7ex\hbox{E}\kern-.125emX}}
\begin{document}

\title{Fusion of Federated Learning and \\Industrial Internet of Things: A Survey
}

\author{\IEEEauthorblockN{Parimala M, Swarna Priya R M, Quoc-Viet Pham, Kapal Dev, \\Praveen Kumar Reddy Maddikunta, Thippa Reddy Gadekallu, and Thien Huynh-The}

\thanks{Parimala M, Swarna Priya R. M, Praveen Kumar Reddy Maddikunta and Thippa Reddy Gadekallu are with the School of Information Technology and Engineering, Vellore Institute of Technology, Vellore 632014, India (e-mail: parimala.m@vit.ac.in, swarnapriya.rm@vit.ac.in, praveenkumarreddy@vit.ac.in, thippareddy.g@vit.ac.in).}

\thanks{Quoc-Viet Pham is with the Research Institute of Computer, Information and Communication, Pusan National University, Busan 46241, Republic of Korea (e-mail: vietpq@pusan.ac.kr).}

\thanks{Kapal Dev with CONNECT Centre, Trinity College Dublin, D02 PN40, Ireland, (e-mail:kapal.dev@ieee.org).}

\thanks{Thien Huynh-The is with the ICT Convergence
Research Center, Kumoh National Institute of Technology, Gyeongsangbuk-do 39177, Republic of Korea (e-mail: thienht@kumoh.ac.kr).}
}
\maketitle

\begin{abstract}
Industrial Internet of Things (IIoT) lays a new paradigm for the concept of Industry 4.0 and paves an insight for new industrial era. Nowadays smart machines and smart factories use machine learning/deep learning based models for incurring intelligence. However, storing and communicating the data to the cloud and end device leads to issues in preserving privacy. In order to address this issue, federated learning (FL) technology is implemented in IIoT by the researchers nowadays to provide safe, accurate, robust and unbiased models. Integrating FL in IIoT ensures that no local sensitive data is exchanged, as the distribution of learning models over the edge devices has become more common with FL. Therefore, only the encrypted notifications and parameters are communicated to the central server.  In this paper, we provide a thorough overview on integrating FL with IIoT in terms of privacy, resource and data management. The survey starts by articulating IIoT characteristics and fundamentals of distributive and FL. The motivation behind integrating IIoT and FL for achieving data privacy preservation and on-device learning are summarized. Then we discuss the potential of using machine learning, deep learning and blockchain techniques for FL in secure IIoT. Further we analyze and summarize the ways to handle the heterogeneous and huge data. Comprehensive background on data and resource management are then presented, followed by applications of IIoT with FL in healthcare and automobile industry. Finally, we shed light on challenges, some possible solutions and potential directions for future research.
\end{abstract}


\begin{table}[h!]
\centering
\caption{Acronyms}
\label{tab:Acc}
\begin{tabular}{ll}
IoT  & Internet of things           \\
IIoT & Industrial Internet of Things \\
DL   & Deep Learning                 \\
ML   & Machine Learning              \\
FL   & Federated Learning            \\
DP   & Differential Privacy           \\
FML  & Federated Machine Learning \\
FDL  & Federated Deep Learning    \\
AI   & Artificial Intelligence    \\
DRL  &  Deep Reinforcement Learning \\
DML  &  Distributed Machine Learning \\
FTM  &   Federated Tensor Mining \\
IoE  & Internet of Everything \\
RFID & Radio-frequency data \\
SDN  & Software Defined Network \\
BF   & Bloom Filter \\
FFBF & Fuzzy Folded BF \\
IDMS & Industrial Data Management System \\
SOA  & Service Oriented Architecture \\
IWNs & Industrial Wireless Networks \\
CCPSA & Cyber Physical System Architecture \\
CPS  & Cyber Physical Systems \\
MTs  & Mobile Terminals \\
MPC  & Multi-party Computation \\
PEFL & Privacy Enhanced Federated Learning \\
PSO  & Particle Swarm Optimization\\
BS   & Base station\\
MBS  & Macro Base Station\\
SBS  & Small Cell Base Station\\
IFL  & Industrial Federated Learning\\
DTWN & Digital Twin Wireless Networks \\
DFRL & Deep Federated Reinforcement Learning \\
DFQL & Deep Federated Q-Learning \\
MAQL & Multi-Agent Deep Q-learning \\
UAVs & Unmanned Aerial Vehicles \\
GFC  & Ground Fusion Centre \\
LSTM & Long-Short Term Memory\\
FIL & Federated Imitation Learning\\

\end{tabular}%
\end{table}

\section{Introduction}
\label{Sec:Introduction}

With an unprecedented increase in the number of Internet of things (IoT) devices and emerging applications, a large amount of traffic is created every day. Such an increase poses a great burden on the Internet network and also demands significant investments for the infrastructure upgrade. However, thanks to the development of big data analytics and artificial intelligence (AI) techniques such as deep learning (DL) and machine learning (ML), the data collected can be effectively exploited for many purposes. From the communications perspective, the last few years has witnessed the emergence of AI applications in various fields. For example, ML is employed to investigate an efficient antenna selection in multi-antenna wireless systems \cite{joung2016machine}, DL is used to handle the computation offloading problem in IoT systems with edge computing \cite{li2018learning}, and deep reinforcement learning (DRL) is used to optimize resource allocation problems at the network edge such as traffic classification, edge caching, network security, and data offloading \cite{luong2019applications}. However, the conventional AI models typically require central processing of data collected from all the users in the network, that is, users should upload their own data to a central server for training the learning model. However, a critical concern with the central learning is data privacy, i.e., some users want to keep track of their local data and do not want to transmit their local data to the central server. 
Training the learning model centrally requires a central cloud with immensely powerful computing capabilities and storage capacities. Meanwhile, recent advances in computing hardware and the proliferation of smart devices in our daily lives have shown that each IoT device can be equipped with reasonable computing and storage levels, which is closely comparable to a desktop computer ten years ago \cite{Pham2019ASurvey_MEC}. Therefore, the standard ML model is not readily applicable to large-scale IoT networks and cannot exploit the availability of distributed computing. 
This calls for a new learning model that leaves the training data distributed across individual IoT devices instead of being centralized.

\begin{figure*}[h!]
	\centering
	\includegraphics[width=0.60\linewidth]{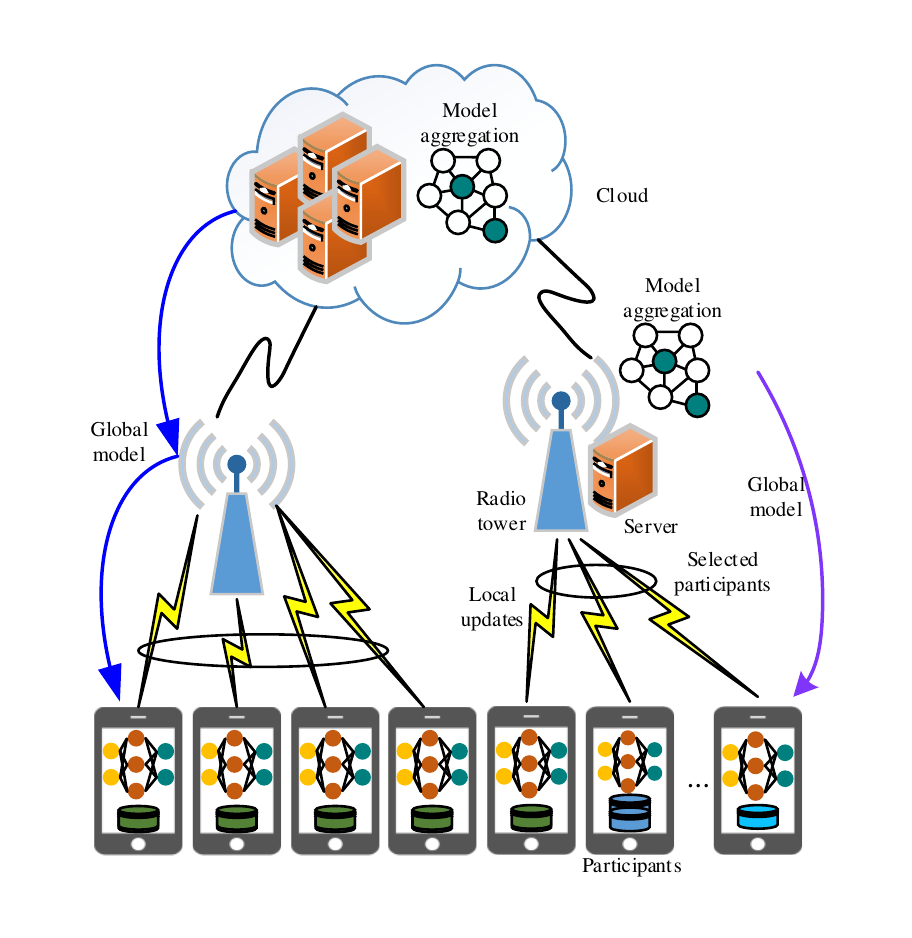}
	\caption{An overview of FL in IoT systems, adapted from \cite{pham2020intelligent}.}
	\label{Fig:FL_Model}
\end{figure*}

Motivated by this issue, Google invented the concept of \emph{federated learning} (FL) for on-device learning and data privacy preservation \cite{konevcny2016federatedlearning}. Using the FL approach, each IoT device can train its model based on locally collected data. 
Local data from IoT devices does not need to be transmitted to the centralized cloud. The centralized cloud only needs to collect the updated local training model from individual users. Thanks to its characteristics, FL has been adopted in many applications, for example, FL for improving Google keyboard suggestions \cite{yang2018applied}, FL for healthcare \cite{xu2019federated}, FL for smart city sensing \cite{jiang2020federated}, and FL for medicine \cite{sheller2020federated}. To pictorially illustrate the concept of FL, an overview of FL in IoT systems is shown in Fig.~\ref{Fig:FL_Model}. In general, each IoT device has its own dataset and the aggregation server can be either located at the network edge or a virtual cloud in the remote cloud computing system \cite{niknam2019federated}. Each FL model has its own pros and cons, depending on various factors. For example, FL with the server at the network edge is suitable for applications requiring low latency, location awareness, and network contextual information \cite{Pham2019ASurvey_MEC} while cloud-based FL is suitable for applications with massive IoT devices over multiple regions and requirements of powerful computing/storage capabilities. 

Although FL has found many benefits in enabling privacy-preserving learning solutions for IoT systems, it also poses significant challenges such as resource management, robustness, security, and incentive mechanisms. These challenges would be more significant and difficult to handle in \emph{industrial IoT} (IIoT), in which much higher levels of security, safety, and reliability are demanded. To enable FL in IIoT applications, a number of research works have been dedicated over the last few years. For example, Liu \emph{et al.} \cite{lu2019blockchain} investigated a solution for IIoT data sharing applications using blockchain and FL and showed that the proposed scheme can enhance security and performance utility without requiring a central server for training. Similar to \cite{lu2019blockchain}, Zhang \emph{et al.} \cite{zhang2020blockchain} employed blockchain and FL to detect device failure in IIoT environments and to secure raw data of IIoT devices. More recently, Liu \emph{et al.} \cite{liu2020deep} developed a communication-efficient FL approach to detect device failures in IIoT through the use of FL and the long-short term memory (LSTM) model. The advantage of this work is that data privacy is preserved via FL and at the same time time-series data in IIoT can be efficiently handled via the LSTM learning model. In spite of the fast development of FL and the rapid emergence of many IIoT applications, there is a lack of a comprehensive survey that provides an overview of FL, IIoT, and applications of FL to IIoT areas. Motivated by this fact, in this work we will be focusing on providing the fundamentals of FL, IIoT, and reviewing state-of-the-art research works on FL for IIoT applications.    

\subsection{State-of-the-Arts and Contributions}
Although there have been a number of surveys focusing on FL and IIoT, these two topics are usually studied separately. With regards to IIoT, the most well-known survey is presented in \cite{da2014internet} with an introduction to IIoT, enabling technologies, major applications, and research outlooks. Similar surveys of IIoT on enabling technologies and applications can be found in \cite{lu2017industry, khan2020industrial}, in which the term "Industry 4.0" is used instead. The roles of big data analytics in IIoT and a thorough classification, including data sources, big data tools, big data techniques, requirements, applications, and analytics types, are presented in \cite{ur2019role}. More recently, a survey on IIoT security and the application of edge/fog computing is presented in \cite{tange2020systematic}.

With regards to FL, several surveys have been conducted over the last few years such as \cite{yang2019federated, li2020federated, lim2020federated, mothukuri2020survey, lyu2020threats, aledhari2020federated, kulkarni2020survey}. The two surveys in \cite{yang2019federated, li2020federated} present an introduction to FL along with general applications and challenges. Lim \emph{et al.} \cite{lim2020federated} firstly provided fundamental features and unique characteristics of FL, and then reviewed three major aspects of FL at mobile edge networks, which are communication costs, resource allocation optimization, and privacy and security. In \cite{mothukuri2020survey, lyu2020threats}, security and privacy issues in FL systems are reviewed. In particular, Mothukuri \emph{et al.} \cite{mothukuri2020survey} tried to address four major questions, including the source of vulnerabilities, the kinds of security threats/attacks, comparison with security and privacy issues in conventional and distributed ML methods, and the defensive techniques. A survey on enabling technologies, network protocols, and promising applications of FL can be found in \cite{aledhari2020federated}. A short survey on personalized models in FL is presented in \cite{kulkarni2020survey}, which concluded that personalized models would enhance the learning performance of individual users in comparison with the global model and local models. We can also find interesting topics about FL in recent magazine articles. For example, the integration of FL into 6G wireless systems is discussed in \cite{Liu2020FederatedL6G}, FL for intelligent fog/edge cloud radio access networks is introduced in \cite{Zhao2020Federated}, a framework of reliable FL is proposed in \cite{kang2020reliable}, FL-based intelligent IoT over controllable communication channels via reconfigurable intelligent surfaces was studied in \cite{Yang2020Federated}, and incentive resource allocation mechanisms using game-theoretic approaches are investigated in \cite{khan2020federated}. A recent survey on the integration of FL and IoT is provided \cite{khan2020federatedSurvey}. This survey also provides a classification of FL-IoT studies, which are cloud server design, miner classification, edge collaboration, federated optimization schemes, incentive design, security and privacy, and global update strategies. 

Different from existing works, in this work, we set to provide a survey on the use of FL for IIoT applications. To the best of our knowledge, this work is the very first attempt on the integration of FL and IIoT. The primary contributions offered by our work can be summarized as follows.
\begin{itemize} 
    \item We present the fundamentals of distributed learning, FL, IIoT architecture and characteristics as well as motivations of integrating FL with IIoT. 
    
    \item We review the state-of-the-arts on FL-IIoT, including FL for enhancing security and privacy in IIoT, data management and resource management in IIoT, and promising application areas such as autonomous industry and healthcare industry. 
    
    \item We discuss a number of challenges associated with current studies on FL-IIoT, and further highlight open research issues that should be efficiently addressed to fully realize FL-IIoT solutions. 
\end{itemize}

\subsection{Paper Organization}
The remaining parts of this paper are organized as follows. In Section~\ref{Sec:Fundamentals}, we present fundamentals of IIoT and FL as well as motivations of FL-IIoT integration. In Section~\ref{sec:FL_SecurityIIoT}, we review the use of FL to provide secure solutions in IIoT. Next, Section~\ref{sec:DataManagement} presents data management and resource allocation in IIoT and discusses the use of FL. Applications of FL for important sectors in IIoT, including autonomous industry, smart healthcare industry, and IIoT security, are reviewed in Section~\ref{sec:Applications}. Challenges and future directions in the use of FL for IIoT are discussed in Section~\ref{sec:Challenges}, and finally we conclude the paper in Section~\ref{sec:Conclusion}. For clarity, the list of frequently used acronyms is summarized in Table~\ref{tab:Acc}.

\section{Industrial IoT and Federated Learning}
\label{Sec:Fundamentals}
\subsection{Fundamentals of Distributed Learning}

The rapid increase in digitization through several advances in information technology applications such as IoT, smart cities, industry 4.0, etc. has led to tremendous growth in the digital data \cite{bhattacharya2020review}. Traditional ML approaches where the ML algorithms are trained using the data stored in a personal computer or at a centralized locations may not work on real-time applications in this era of big data due to the limitations in scalability and time complexity. If the training data is too large, the ML algorithms may not be trained properly due to the memory and computational constraints. To address this issue, distributed ML (DML) has been introduced. In DML, the ML algorithms can be trained on the subsets of the data which are distributed across several nodes \cite{wu2019stability,zhang2020lagc}.

Distribution of the training data in DML can be performed by following two approaches. The first approach is horizontal fragmentation, where the instances or rows in the dataset are divided into several fragments and each fragment is stored in a node in the distributed network. The second approach is vertical fragmentation, where a subset of attributes are distributed across several nodes in the distributed network. The ML algorithms can then be executed in parallel in each of the node in the distributed network, which reduces the training time of the ML algorithms significantly, thus making the ML algorithms scalable and addressing the limitation of memory requirements for training the ML algorithms. Some applications generating large amount of data in real-time where DML is extensively used are healthcare applications, IoT, smart city based applications, communications, etc. \cite{du2020approximate,rossi2020distributed,dev2020triage,raja2020ai,bogowicz2020privacy}. DML also plays a huge role in big data analytics \cite{xu2020distributed,reddy2020analysis}.

Even though DML has \textcolor{black}{enormous} potential in several applications, it has its own limitations. Some of the limitations of the DML are privacy, security, platform dependency, lack of standard measures to evaluate the DML algorithms \cite{huang2019dp}. 
The following subsection introduces the concept of FL, how it addresses the issue of privacy and security in \textcolor{black}{conventional ML and DML.}

\subsection{Fundamentals of Federated Learning}

FL is a specific category of DML where the ML models are trained on the data located at the decentralized devices such as mobile phones and other smart devices \cite{yang2019federated}. The devices participating in the FL need not exchange their data samples, that ensures the privacy and security \cite{li2020federated}. The underlying principle in FL is that the ML models are trained on local data, then the parameters of the ML models are exchanged between the local nodes (devices) at regular intervals that enables the creation of a global ML model. The training data remains on the individual devices. The global ML model resides at a server. Once all the devices send their models to the server, a \textcolor{black}{combined model} is created by averaging the parameters of the individual models. In this way, the individual devices learn collaboratively from a shared model. Some of the advantages of FL over traditional ML models are that FL ensures \textcolor{black}{data privacy}, reduced latency, reduced power consumption \cite{xu2019verifynet}.  \textcolor{black}{Not only the FL ensures privacy of sensitive information of users, but also FL can deliver personalized ML models to the users with enhanced user experience \cite{kang2020reliable}.}

The basic difference between FL and DML is that the DML concentrates mainly \textcolor{black}{on} parallelization of the computing power whereas FL concentrates on training the datasets that are generated locally on devices such as mobile phones. The DML assumes that the datasets are distributed identically (in size) across several nodes in the architecture. Whereas, in FL, the size of the datasets at different nodes may be heterogeneous \cite{sattler2019robust}. \textcolor{black}{Also,} the nodes in the DML are typically data centers that can communicate with latest communication technologies and they have access to powerful computational resources. On the other hand, the nodes in the FL are typically devices that may be \textcolor{black}{battery-dependant} and also work on communication media like \textcolor{black}{Wi-Fi and cellular networks} that makes them unreliable and drop out frequently from the network.

\subsection{Characteristics of IIoT}
IIoT is collection of people, sensors, machines, and computers that enable intelligent industrial operations with the help of ML and analytics \cite{mathur2020overview}. IIoT enables intelligent devices and networked sensors to collect the data from the manufacturing plants and make use of predictions through AI and automate the decision making \cite{farivar2019artificial, liao2019learning}. IIoT offers \textcolor{black}{a} revival of industries \textcolor{black}{that} have been lately struggling due to several factors including shortage of skills from personnel. IIoT offers several benefits because of their minimal or nil dependency on human intervention. IIoT systems can make intelligent decisions by automatically learning from the data generated through the sensors. IIoT leads to smart machines that can achieve highest levels of accuracy that was not possible earlier because of human intervention. This disruptive technology has huge potential to transform the way the products are manufactured and delivered, make factories smart and efficient, protect front line workers by providing better security in difficult conditions, reduce human errors, increase efficiency, thus, in turn, save huge amount of money \cite{vinayakumar2020visualized,zheng2020secure}.

Many sectors and applications such as manufacturing, transportation system, oil and gas, healthcare, agriculture, smart cities, energy are benefited by IIoT. One of the main benefits of IIoT is its ability to increase efficiency in operations through prediction. For instance, the sensors in a machine can automatically pinpoint specifically where the trouble is when it goes down and place a service request automatically. The sensors can predict the likelihood of a breakdown of machines before it happens by automatically sensing the sounds of a machine, temperature, vibrations, and other factors by using data analytics and ML. Hence, predictive maintenance reduces the idle time of a machine and also reduces the damages caused by faulty machines by fixing the issues before they are escalated \cite{huang2019real}. Another important advantage of IIoT is location tracking of equipment and tools. In many industries, the workers spend significant amount of time \textcolor{black}{searching} for tools, finished goods in an inventory. IIoT sensors saves lot of time by providing location services, thereby making it impossible to lose equipment, goods\cite{liu2020reliability}. Until recently, manufacturing companies used to spend huge capital on purchasing and maintenance of equipment. But with the advent and advancements in IIoT, the industries can lease the equipment from the vendors. The owners can remotely monitor these equipment through the sensors, thereby delivering upgrades, repairs, and maintenance remotely. This will save both time and capital for industries\cite{aceto2019survey}.   
To summarize, the main benefits of IIoT are 1) reduced waste, 2) reduced maintenance costs, 3) energy savings, 4) workforce productivity gains, 5) improved service, and 6) revolutionary products and services.

    
    
    
    

\subsection{\textcolor{black}{Motivations of Integrating Federated Learning with IIoT}}

The exponential growth in the IIoT applications is being hindered by several issues such as security, privacy, communication cost, etc. Integration of FL with IIoT has the potential to solve many of issues discussed above. \textcolor{black}{The main motivations behind integrating FL with IIoT applications are summarized as follows:}

\begin{itemize}
    \item \textbf{Security and Data Privacy Preservation}: In order to get valuable insights and patterns from the data generated by IoT devices in IIoT applications, ML \textcolor{black}{and} DL algorithms are used. To understand the complex patterns from the data generated, these algorithms are trained regularly by large datasets collected from different industries/locations. But moving these datasets to a central location for training the algorithms exposes the sensitive data to potential hackers, \textcolor{black}{and} intruders\cite{cmc.2021.013852}. The datasets need not be transferred to a central location for training the algorithms when FL is used. Hence it secures the sensitive data from potential attackers. This also preserves the privacy of sensitive data.
    
    \item \textbf{Reduced Communication Cost}:
    The sensors in IIoT applications generate large volumes of data. This data is stored in the edge devices for analytics and then moved to the centralized locations such as \textcolor{black}{the} cloud. Transferring huge volumes of data incurs significant communication cost. With FL, the entire data need not be transferred to the cloud. Instead, only the summarized results after application of the the ML/DL models is transferred to the cloud. This reduces the communication cost for transferring the data to the cloud.  
    
    \item \textbf{Improved Performance of the Network}:
    A ML/DL model is implemented on the data generated from IIoT \textcolor{black}{applications} is stored in an edge device and only \textcolor{black}{the} summary of the results is transferred to the central location when FL is used. Hence, the traffic in the network is reduced considerably. With reduced network traffic, the overall performance of the network \textcolor{black}{can be} significantly improved.
    
    \item \textbf{Scalability}:
    Integration of FL with IIoT enables the DL algorithms to scale their learning as they need not be trained on large volumes of data generated locally. The central algorithm is trained only on summarized results from the individual edge devices. This facilitates \textcolor{black}{a} significant scalability of the central learning algorithms and train from the data generated at several edge/local devices.
\end{itemize}

\section{Federated Learning for security in IIoT }
\label{sec:FL_SecurityIIoT}
FL protects user data by training the model without transferring the data from client to the main server. The scope of security mechanism can be further enhanced while sharing the model updates and integrating the data generated from similar devices from various industries. In this section, we discuss about providing different security mechanisms by integrating FL with ML and DL models in IIoT framework. FL with blockchain in IIoT is also presented in this section.
 
\subsection{Federated Machine Learning for Secure IIoT}

The amount of data generated from a single industry will not be suitable for applying ML techniques \cite{li2018deep}. Generally, many industries would have similar IoT surveillance systems \cite{da2014internet} and so the data generated can be federated from all the devices that could lead to increase in the accuracy of ML algorithms. However, the  key  concern  is  the  data  security  \cite{huang2019towards} provided  during leveraging  of  data. Federated \textcolor{black}{ML} (FML) has gained a lot of attention because the way it handles the  privacy by decentralizing the data generated at the end user device and aggregation of ML models at a centralized server.  
The two types of attacks in IIoT namely eavesdroppers and hackers were discussed in \cite{kong2019federated}.
\begin{enumerate}
    \item \emph{Eavesdroppers}: Attack happens when the data is transferred through communication channels. In this case, it may happen between IoT node and IoT sink, then from IoT sink to centralized server communication channels.
    \item \emph{Hackers}: Attack that happens in the centralized server.  A stalker resides in the server and is capable of obtaining actual locations of a user.
\end{enumerate}

As discussed in \cite{kong2019federated}, initially the data is generated from IoT devices from every smart industry sent to the IoT sink.  The  sink is a repository for collecting the data from different IoT nodes in  the  industry  through  wired  and  wireless  communications and it also encrypts  the  data which is sent  to  the  centralized server. Then the server collects the information from multiple IoT sinks and federates the data.  Finally, the smart industry decrypts the knowledge into an understandable format.  In  this  case,  both  eavesdroppers and  hackers  attacks  are  not  possible,  as  encrypted  data  is present in both the cases.

The  main  intention  of  FL  is  to  collectively  learn  the global  model  without  sacrificing  the  privacy of data. But sharing model updates to the server during training process  may  lead  to  an  issue  of  leaking  sensitive  data  like user personal information. In order to preserve privacy in ML models, number of methods are employed in FL.
\begin{itemize}
    \item \emph{Differential Privacy (DP)}: DP defines the amount of information about the data that can be available for third party analysis \cite{arachchige2020trustworthy}. Information under DP can be categorized as general information which has the information about the  entire  population and the other type is  private information that has the  personal private data. Some of the properties of DP framework to provide sensitive personal information and privacy protection includes quantification of privacy loss, composition,  group  privacy  and  closure  under post-processing.   

    \item \emph{Homomorphic Encryption:} Computation and analysis is done based on encryption technique so that the attacker finds very hard to find the original information \cite { kuang2015secure}.

    \item \emph{Secure Multiparty Computation:} It is a model where multiple parties collaboratively compute without leaking any information to the third parties \cite{raja2020ai}.
\end{itemize}

To summarize, \cite{arachchige2020trustworthy} DP does not guarantee complete data security \textcolor{black}{because if the algorithm wrongly classifies a private information as general information, then the algorithm fails to protect the data.} In spite of all these drawbacks, DP has proved to be more efficient privacy preserving approach compared to the traditional centralized method. 

\begin{figure}[t]
	\centering
	\includegraphics[width=1.0\linewidth]{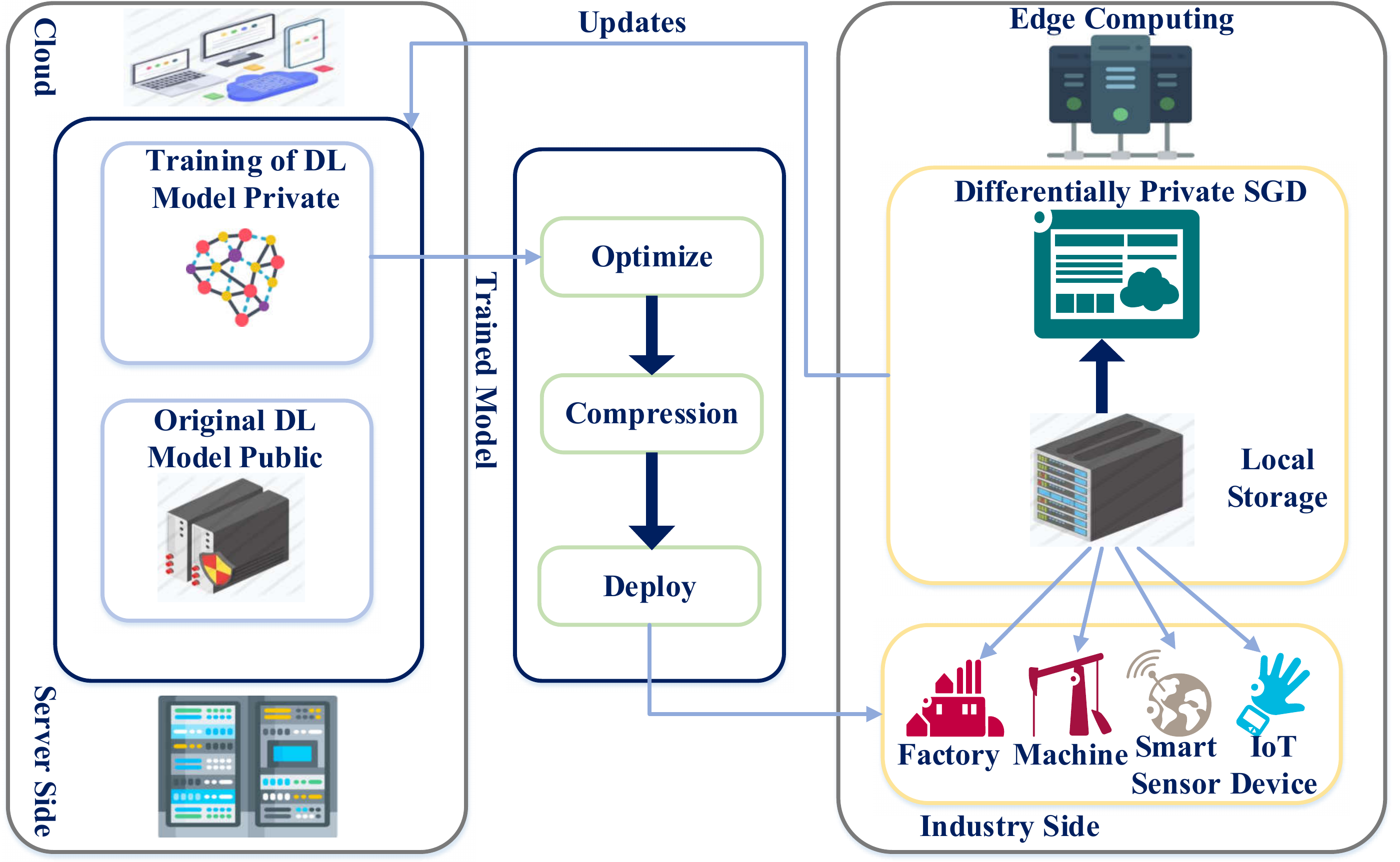}
	\caption{\textcolor{black}{Federated DL in IIoT.}}
	\label{Fed2}
\end{figure}

\subsection{Federated Deep Learning in IIoT}
Recently, integration of DL models with IoT and edge devices have become more popular which  provide  real  time analytics with limited resources \cite{huynh2020mcnet}. So, federated DL (FDL) enables Industry 4.0 companies to integrate DL in IoT devices and provides secure framework using FL as shown in Fig.~\ref{Fed2}. DL is computationally expensive which requires costly resources and framework. So decentralising DL models is a multi-dimensional issue which requires new technologies framework to integrate DL with edge computing and IIoT. The main goal of FDL is to provide IIoT with advanced capabilities using optimized DL which would transform the Industry 4.0 factories into smart factories. Some of the parameters that are required to build the FDL model in IIoT are discussed below. 
\begin{enumerate}
    \item \textbf{FDL Model}: A FDL model can be implemented on both client and server side. In client side, private networks are defined which have the DL model tuned and optimized from the general model present in the cloud. Optimized and tuned models are then deployed on the client side, where the model is trained with locally generated data from the end device. Finally the end device contains highly quantized and compressed FDL model. On the server side, model present in cloud is continuously updated by integrating differentially gradients from each private network. Each local DL network takes its turn to continuously upload and download the currently updated gradients to the cloud model. So, distributed selective stochastic gradient descent approach is presented in \cite{shokri2015privacy} which can be applied in cloud model to frequently update local private model. The first decentralised model named “Model Chain” uses blockchain technology \cite{deepa2020survey, hakak2020recent} to enable privacy preserving in the data transfer. In addition to that, asynchronous stochastic gradient descent can also be used when a single model can be trained in parallel among all the devices, aggregated and processed.

    \item \textbf{FDL Communication and Networking}: The main advantage of using FDL is to run DL models in IoT devices and to involve model in decision making process. This kind of decentralizing DL process enhance robustness, operational efficiency and reliability of IoT devices. FDL provides two types of communication, namely, intra-communication channel and inter-communication channel. Former transmits the data among all the tiers in the framework. FDL communicates between the IoT and cloud tier where the optimized model in the cloud is deployed to the end device. However, the security and privacy must be maintained in the FDL during communication. In inter communication channel, the components in each layer communicates with each other in three different ways such as cloud, edge and end device. The main aim of FDL is to minimize intra-communication and maximize inter-communication which would significantly reduce the cost of communication.
    \item \textbf{FDL privacy and security}: Lim \emph{et al.} \cite{lim2020federated} discussed about various security threats while FDL shares DL models from cloud to end devices. Anyway, FDL builds DL models that do not expose information about the data to the cloud. Security issue on the server side includes sharing of DL models on the cloud that leads to confidentiality and security risk. Security issue on the client side is done by encrypting the data during the training process before sending it to the cloud server. Some mechanisms namely DP and homomorphic encryption technique controls the amount of data to be shared on the cloud.
    \item \textbf{FDL Optimization}: As the end devices have limited memory and computational requirements, DL models must be optimized so that they can be deployed on IoT or end devices efficiently. In terms hardware optimization, GPU offers low power computation that decreases computation time. FPGA and Google’s Tensor Processing Unit \cite{wang2019benchmarking} are other DL devices that boosts the processing of DL network. In terms of memory optimization, algorithms like shared memory allocation algorithm can be used for DL models. Dynamic scheduling \cite{cho2012benefits} is one of the key process used for optimizing the performance on the cloud server. 
\end{enumerate}

To summarize, the difference between the above two sections FML and FDL is presented in Table \ref {tab2}.  Recently, researchers in \cite{zhang2019deeppar,wang2020federated,liu2020privacy,chen2019communication} proposed DL models in FL for IIoT networks in various application such as automobile and mobile network, traffic network and in image processing applications.
\begin{table}[t]
\centering
\caption{\textcolor{black}{Difference between FDL and FML.}}
\label{tab2}
\begin{tabular}{|p{1.8cm}|p{2.85cm}|p{2.85cm}|}
\hline
\multicolumn{1}{|c|}{\textbf{Features}} & \multicolumn{1}{c|}{\textbf{FDL}}                & \multicolumn{1}{c|}{\textbf{FML}}                   \\ \hline
Communication                           & Communication between cloud and IoT is possible  & Communication between cloud and IoT is not possible \\ \hline
Privacy      & Privacy is less than FML                & Privacy is more               \\ \hline
Data storage & Private DL model is maintained in cloud & No data resides in cloud      \\ \hline
Updates                                 & Gradient DL models updates are sent to the cloud & Statistical updates are sent to the cloud           \\ \hline
Devices connected with Cloud            & IoT devices are directly connected with cloud    & Edge devices act as a buffer between IoT and Cloud  \\ \hline
Model access & Model access is   private               & Model access is public        \\ \hline
Data control & Data control is centralized             & Data control is decentralized \\ \hline
Optimization & Optimization is   required              & It is not required            \\ \hline
\end{tabular}%
\end{table}

\subsection{Blockchain in IIoT using FL}
Generally, FL uses a central server to aggregate all the updates of ML and DL models and communicates the aggregated updates to the global model. Security breach can happen when the updates are transferred to the server or when the global model is sent to the clients. So to overcome this security issue, the blockchain concept can be applied to store the model updates in immutable form. \textcolor{black}{Yin \emph{et al.}} in \cite{yin2020fdc} presented a secured model that uses blockchain as a decentralized framework to replace the traditional central servers. Also, it stores the historical information as tamper resistant data. Lu \emph{et al.} \cite{lu2019blockchain} investigated some security issues in node, block and framework which are designed using blockchain in FL.

\begin{itemize}
     \item \emph{Node security:} In terms of node security, blockchain ensures the node permission control using alliance chain and data security part is implemented using the chain of nodes. Each node is created when the parameters are updated or model is modified. Each validated update is stored in the form of chain which is untampered or immutable. Committee consensus mechanism can also be used to validate the node to verify whether they are malicious or not. Committee allows the good node to send the updates and change the global model for efficient training. If it is a malicious node updates, then the FL will ignore the updates which would affect in designing of global model.
     
    \item \emph{Block security:} It is a vital process in blockchain to implement the consensus for each block. Executing and broadcasting the consensus is highly time consuming task, so it is inevitable to reduce the consensus cost. 
    
    \item \emph{Model Security:} The data is secured in FL as there is no transfer or exchange of information between the client and the server. But the global model constructed in the centralised server can be hacked by the unauthorized person. Model trained in the central server is more exposed to attack compared to model stored in local server. So the model build and trained should be preserved at both the ends. \end{itemize}

\begin{figure}[t]
	\centering
	\includegraphics[width=\linewidth]{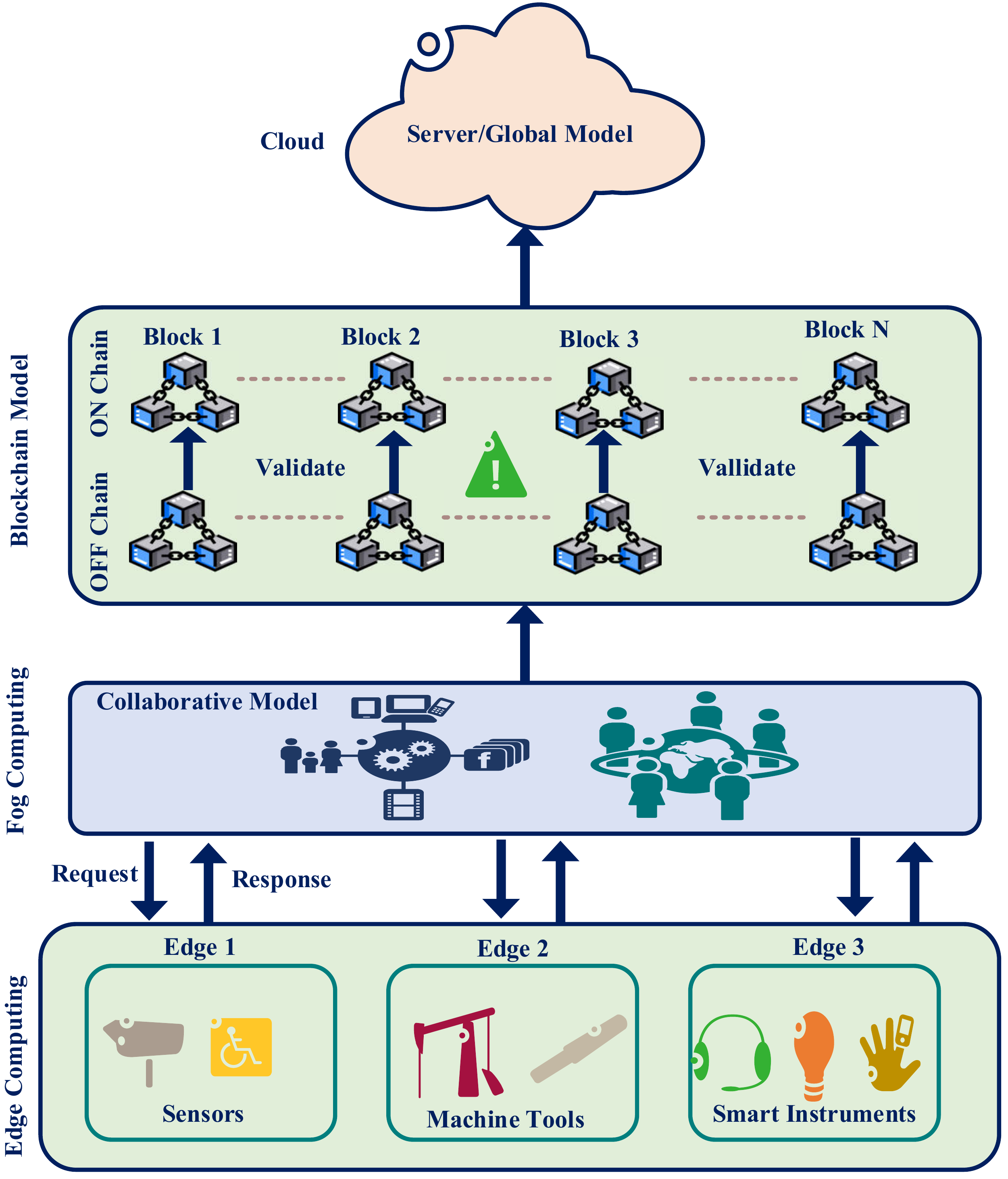}
	\caption{\textcolor{black}{Blockchain in IIoT using FL.}}
	\label{Fed1}
\end{figure}

The workflow of FL with blockchain is shown in Fig.~\ref{Fed1}. IIoT devices are designed in such a way that \textcolor{black}{they} train the model with local data. After training, \textcolor{black}{they} send the local updates to the edge node where all the local updates are consolidated. Edge nodes in turn send aggregated local updates to the fog node where the updates from all the edge node are aggregated. Then, the updates are converted into a block, which is verified based on the smart contract. The validated block is joined with onchain which is tamper-proof and secured. Finally, all the local updates are received by the centralised server to train the global model. Then the updates of the global model is sent to all the edge node to update on all the end devices connected to the edge node.

The framework must be scalable for real world applications that can adapt to different kind of scenarios. In recent works \cite{kim2019blockchained,qu2020decentralized, pokhrel2020federated}, researchers have applied blockchain in image processing and vehicular networks using FL in IIoT framework. The works in \cite{li2020crowdsfl,hua2020blockchain,sharma2020blockchain} designed a FL based blockchain framework for health industry, railway network and defence organization.   \textcolor{black}{Generally, privacy preserving mechanism in FL using IIoT will be applied on end to edge device communication and from edge to collaborative model.} Even in worst case also only the model updates can be tampered between end devices to collaborative engine. Recent studies of FL in ML, DL and blockchain in IIoT are summarized in Table \ref {tab1}.

\begin{table*}[!ht]
\centering
\caption{Overview of recent studies on FL with secure IIoT.}
\label{tab1}
\resizebox{\textwidth}{!}{%
\begin{tabular}{|l|p{1.3cm}|p{1.40cm}|p{2cm}|p{3.5cm}|p{5cm}|p{3.8cm}|}
\hline
Ref&
  Framework &
  Algorithm &
  Application &
  Techniques &
  Contribution Inference &
  Limitations challenges \\ \hline
\cite{kong2019federated} &
  \multirow{18}{*}{\makecell{ FMLin IIoT}} &
  FTM &
  Smart   factories that belong to same vertical industry &
  FTM   with homomorphic encryption which leverages the linear transformation &
  $\circ$ Every factory has to share its ciphertext data
  
  $\circ$ Mines the same knowledge from different   factories   
  
  $\circ$ Defend the   attacks from distributed and centralized hackers
  &
  $\circ$ Challenge to   maintain balance between ML and data security
  
  $\circ$ FTM needs the   centralised server to operate the federated mining
  \\ \cline{1-1} \cline{3-7} 
\cite{arachchige2020trustworthy} &
   &
  Primalchain &
  NLP   and speech recognitiono &
  DP &
  $\circ$ Uses   FML to generate global representation of distributed knowledge in IIoT   environment
  
  $\circ$ Blends   DP, FML and blockchain &
  $\circ$ Large scale ML latency is more
  
  $\circ$ Efficiency can   be improved \\ \cline{1-1} \cline{3-7}   
\cite{guo2011tensor} &
   &
  Tensor   Ridge Regression &
  Image   processing &
  Parallel   factor decomposition &
  $\circ$ Projections of   input tensor to more than one directions
  
  $\circ$ Multilinear   mapping from tensorial input space to continuous output space &
  $\circ$ High   dimensional datasets leads to overtraining, high computational complexity and   large memory requirement \\ \cline{1-1} \cline{3-7} 
\cite{lai2014sparse} &
   &
  STA &
  Computer   vision and pattern recognition &
  Tensor   Alignment Technique &
  Unsupervised   tensor feature extraction
  
  Enhances   the robustness in the alignment step
  
  $\circ$ Works good for high dimensional space
  &
  Sparse   projection learning methods are not investigates for tensor recognition
  
  Not efficient   for high dimensional space \\ \cline{1-1} \cline{3-7} 
\cite{feng2018privacy} &
   &
  CPSS &
  Wireless IoT &
  $\circ$ Privacy   preserving high order singular value
  
  $\circ$ Decomposition   based criterion &
  $\circ$ Model allows users to utilize the storage and computing resources of cloud and fog without leaking any sensitive information &
  As   it uses encrypted data, cost is very high \\ \hline
\cite{zhang2019deeppar} &
  \multirow{18}{*}{\makecell{FDL in IIoT}} &
  \begin{tabular}[c]{@{}l@{}}DeepPAR\\    \\ DeepDPA\end{tabular} &
  Smart   factories &
  $\circ$ Proxy   re-encryption
  
  $\circ$ Group dynamic   key management &
  $\circ$ Protects each   participants input privacy while preserving dynamic update
  
  $\circ$ Guarantees   secrecy in group participants
  &
  Still   security mechanism at edge device can be improved \\ \cline{1-1} \cline{3-7} 
  
\cite{wang2020federated} &
   &
  Double   Deep Q-network &
  Mobile   networks &
  Decentralised   cooperate edge caching
  
  Markov   decision process &
  $\circ$ Federated   reinforcement learning deep framework enables fast training and decouple the   learning process from cloud
  
  $\circ$ Reduces   performance loss and average delay&
  $\circ$ Decentralaized   loss can be maximized \\ \cline{1-1} \cline{3-7} 
\cite{liu2020privacy} &
   &
  FedGRU &
  Traffic   network &
  $\circ$ parameter   aggregation
  
 $\circ$ Federated   averaging algorithm
 
 $\circ$ Joint announcement   protocol &
  $\circ$ Random   subsampling for participating organization is used to reduce the   communication overhead
  
  $\circ$ Captures the   spatiotemporal correlation of traffic flow data &
  Spatial   temporal dependency can be better captured by applying graph convolutional   network \\ \cline{1-1} \cline{3-7} 
\cite{chen2019communication} &
   &
  ASTW-FedAVG &
  Image   Processing &
  $\circ$ Asynchronous   strategy
  
  $\circ$ Weighted   aggregation &
  $\circ$ To reduce the   communication between the client and server 
  
  $\circ$ Asynchronous   strategy is used to aggregate and update the parameters
  
  $\circ$ Previously   trained local models are aggregated to enhance the performance &
  $\circ$ Only server are   involved in evolving the local model
  
  $\circ$ Communication   cost can be reduced \\ \hline
 \cite{yin2020fdc} &
   &
  Federated collaboration framework &
  Wearable sensor data &
  Private and public data center
  
  blockchain &
  $\circ$ Framework for collaborating multiparty computation is designed
  
  $\circ$ Blockchain is   used to overcome security issues &
  Large scale   multiparty secure collaboration of IoT data is challenging \\ \cline{1-1} \cline{3-7} 
\cite{kim2019blockchained} &
  \multirow{28}{*}{\makecell{Federated \\Blockchain}} &
  BlockFL &
  Wireless   Communication Network &
 $\circ$  On device ML
  
  $\circ$ Smart Consensus   mechanism &
  $\circ$ Verified local   updates are generated as block and added to blockchain
  
  $\circ$ Each device   generate global model from the updated new block &
  Proof-of-work can be hacked by the malicious miners \\ \cline{1-1} \cline{3-7} 
\cite{qu2020decentralized} &
   &
  FL-Block &
  Image   processing &
  $\circ$ Decentralized privacy
  
  $\circ$ Poisoning attack &
   $\circ$ Allow local updates to exchange with global blockchain model
  
  $\circ$ As the central authority is replaced by blockchain, privacy is achieved more
  &
  Tradeoff   between privacy protection and efficiency should be optmized \\ \cline{1-1} \cline{3-7} 
\cite{pokhrel2020federated}&
   &
  BFL &
  Vehicular   network &
  $\circ$ Renewal reward approach
  
  $\circ$ VML &
   $\circ$ Provide   efficient communication of autonomous vehicles
  
 $\circ$ It forms   representative groups from the overall vehicles &
  Choice   of block updates will limit the increase in overall delay \\ \cline{1-1} \cline{3-7} 
  
  
  
  
\cite{li2020crowdsfl} &
   &
  CrowdSFL &
  Health   Industry &
  Crowd sourcing   encryption algorithm &
  $\circ$ Data privacy is   controlled by smart contracts
  
  $\circ$ The updates  submitted by the local device cannot be   leaked as it enables the decentralised blockchain platform &
 $\circ$ Reward   distribution mechanism is not defined
  
 $\circ$ Model can be   optimized \\ \cline{1-1} \cline{3-7} 
\cite{hua2020blockchain} &
   &
  SVM   and Deep network &
  Railway   network &
  SVM based on   mixed kernel function and multi class model &
  $\circ$ Distributed   asynchronous collaborative ML is designed without involving the central   server
  
  $\circ$ Smart contract   involves all participant and generates new block &
  Accuracy and   efficiency of the prediction can still be maximized \\ \cline{1-1} \cline{3-7} 
\cite{sharma2020blockchain} &
   &
  IoBT &
  Social   Security through defence organization &
 $\circ$ Multilayered   model
  
  $\circ$ Defence led   combat &
  $\circ$ Distributed   computing defence framework for sustainable society
  
  $\circ$ Optimize the   data trained in local nodes &
  Participation of   local node in training process can be rewarded based on their contribution \\ \hline
\end{tabular}%
}
\end{table*}

\section{Data Management and Resource Management }
\label{sec:DataManagement}

Internet of everything (IoE) is a form of the network, including physical, digital and virtual objects \cite{rm2020load}. These entire groups of objects are networked together to serve the users' requirements. The requirements are served by communicating with each other and these are performed over the  Internet.  The data generated by virtual objects as well as physical objects are heterogeneous and huge in volume such as logs in their SCADA systems\cite{babayigit2019iiot}, data from various \textcolor{black}{business-oriented} applications\cite{leminen2020industrial}, radio-frequency data (RFID) \cite{mathur2020overview}, data from wearable devices\cite{munirathinam2020industry}, data from sensors\cite{reddy2020deep}, real-time web data\cite{parimalaspatiotemporal}, media data and location data\cite{mathur2020overview}. So the available traditional data management techniques fail to handle such \textcolor{black}{heterogeneous and huge data}. There are \textcolor{black}{a variety} of distinct solutions, such as architecture-based solutions, middle-ware based solutions \textcolor{black}{and indexing solutions.} Though traditional ML had ruled the various intelligent smart industrial applications, most of the IIoT applications are not widely deployed in real-time. The reason for the gap in real-time implementation is due to sensitivity of the data and hesitation of the industry in sharing their sensitive information over the Internet and also due to resource constraints in handling the communication cost, throughput and latency. FL can be utilized to overcome these challenges as the technique is specifically designed to maintain data privacy by avoiding global sharing of data.  This section investigates the contributions of researchers in the field of FL for data management and resource management. 

\subsection{Data Storage in IIoT}

The world has now stepped into the era of IoE. IoT is a subset of IoE and IIoT is a subset of IoT. As we move further into this era, the data produced and gathered goes beyond tetra-bytes of data. This data is structured, semi-structured and, in some cases, unstructured. This huge volume of data produced by the IIoT environment is of great value and can be used to extract knowledge about the manufacturing unit or maintenance unit or any unit in an industrial environment so that any kind of abnormality can be prevented in the near future. But storing such huge data and processing them to extract knowledge and reacting at the right time is the biggest challenge now. The technologies, namely fog computing, cloud computing, and edge computing have given the researchers an opportunity to solve the issues with respect to data storage \cite{pham2020coalitional}.

In \cite{fu2018secure}, \textcolor{black}{Fu \emph{et al.}} proposed a framework for processing the IIoT data. Almost all the data life cycle phases are integrated, starting from pre-processing to the retrieval of the requested data stored or archived using fog and cloud computing. The framework was designed to have entities, namely edge server, proxy server, and cloud server for taking care of data storage and processing. The time-sensitive data generated by the IIoT data sources is received by the edge servers where the data is processed at a fundamental level. If the data in the edge server is needed for future usage, then those are transmitted to the proxy server where the quality of the data is improvised by pre-processing, transforming into structured data so that they can be stored in the cloud server as historical data for further predictive analysis.  

Borylo \emph{et al.} \cite{borylo2016energy} proposed and presented a scheme named dynamic resource provision by combining the cloud and edge computing schemes for the purpose of energy-awareness. In this scheme, the major limitation was that the amount of energy consumed was high. In order to overcome this limitation, \textcolor{black}{Kaur \emph{et al.}} \cite{kaur2018edge} designed an architecture consisting of three different layers, namely cloud layer, edge layer and software-defined network (SDN) layer. The SDN layer plays a major role in handling network congestion and hence minimizes energy consumption. 

In \cite{singh2018fuzzy}, \textcolor{black}{Singh \emph{et al.}} proposed a data storage method using bloom filter (BF) and fuzzy logic, namely Fuzzy Folded BF (FFBF). Two different BFs are compressed into a single BF by using fuzzy operations. The major advantage of this compression was that in a single BF, a large volume of data elements was stored. The rate at which the data decayed was reduced so that the data streamed were residing for a longer time. The computational cost was reduced due to the usage of the double hashing technique.

An integrated framework, namely IoT based Industrial Data Management System (IDMS), for an IIoT environment was proposed by \textcolor{black}{Saqlain \emph{et al.}} \cite{saqlain2019framework} for handling and managing the huge industrial real-time data acquired from a smart manufacturing unit. This framework tried to monitor this huge real-time data online and hence make wise decisions. For the purpose of handling emergency events, traditional communication protocols were used. The middle-ware, which was designed in the IDMS, provided a Service Oriented Architecture (SOA). The raw data was transformed into a structured format by the framework so that it can be archived in the cloud server where useful information and knowledge can be extracted.  

In \cite{anton2019highly}, \textcolor{black}{Anton \emph{et al.}} presented a model that can smartly collect, aggregate, and analyze the data collected from heterogeneous sources. The model consists of three levels. The base level, that is the first level, is responsible for collecting singular packets which are correlated along with various logs and the collected packets are analyzed here. The second level is responsible for aggregating all the packets that flow in the network and the information is extracted based on the application. This level is more application-specific. The third level which is the highest level in the aggregation model, aggregates all the raw data and meta-information related to each application into a connected graph which gives an overall picture of the flow of packets in the network.

An IIoT based intelligent control and management system was designed by \textcolor{black}{Du \emph{et al.}} in \cite{du2018iiot} for the purpose of the motorcycle endurance test. The architecture designed in this system is made of four layers, namely the executive layer for the sake of data acquisition, the cognitive layer which takes care of the business logic, the network layer which acts as the communication layer and finally, the control layer. The system provides an interface for establishing a relationship between the protocols and the other external parameters. The design enables easy management, up-gradation, extension and compatibility.

\textcolor{black}{Wan \emph{et al.}} \cite{wan2016software} analyzed the general architecture of IIoT that includes the physical layer comprising of the devices, industrial wireless networks (IWNs) dealing with protocols, industrial cloud for storage and smart terminals for better interaction. As a result of this analysis, this work proposed a software-defined based IIoT architecture consisting of three layers named physical infrastructure dealing with the implementation as well as production, control layer taking care of the interaction between physical and application layer and finally, the application layer for innovative applications. They also assessed the performance of this architecture using a test-bed. The results showed an improvement in the utilization of the equipment and also reduced the consumption of energy.

\textcolor{black}{Shu \emph{et al.}} \cite{shu2016cloud} proposed a novel architecture termed as cloud integrated cyber physical system architecture (CCPSA) which comprises of three domains: one domain which focuses on the network and their communications, other-oriented towards the cloud platform dealing with the control and computational aspects and the final domain is a data-centric CPS. This work also focused on providing solutions for challenges like resource management, scheduling aspects of resources. \textcolor{black}{Liu \emph{et al.}} \cite{liu2018blockchain} proposed a methodology for acquiring and collecting data in an efficient manner so that no data is missed due to the limited energy in the smart terminals using Ethereum blockchain and DRL. The DRL increases the amount of data collected at a time by the mobile terminals (MTs) and the blockchain enables higher security.

To summarize, most of the available data acquisition and data storage techniques as well as frameworks focus on specific issues, but acquiring and storing such huge heterogeneous industrial data is a challenging task. Also the data collected from various factories distributed at different floors and locations can be of different formats and volume\cite{coronado2018part}. Now, most of the industries are shifting towards digitization for achieving their goals. Few challenges that can be listed out are data dynamicity, data visualization, storing industrial data. To meet these challenges a well-versed data management framework has to be devised that can efficiently handle the huge data. 

\subsection{Federated Learning for Data Management}

The IIoT era has enabled the heavy flow of data packets over the Internet, which has increased the traffic. Moreover, recent advancements have made it easier for the industrial giants to gather huge volumes of data from devices located in different geographical locations. Though huge data is collected, most of the industrial applications lose their value in the market as the users are not satisfied. The applications need to process the huge data for making any decision by transferring them to a centralized server or location, which requires high bandwidth, but they fail because of a huge delay in the transportation of data from heterogeneous devices. To overcome this, few heterogeneous architectural based solutions were provided like usage of edge computing, fog computing, etc. But these solutions still have challenges like security, privacy, etc. The major issue with these kinds of architecture-based solutions is that most of the decisions are made locally, neglecting the global data, model and knowledge. This issue could be solved by the usage of a technology proposed by Google named FL \cite{aledhari2020federated}. This section discusses the various approaches coined by various researchers for handling the influence of local knowledge over the decisions. 

\textcolor{black}{Wang \emph{et al.}} \cite{wang2019edge} proposed a framework named "In-Edge AI" which imparted intelligence to the mobile edge network using DRL and FL. This framework is designed to reduce the volume of data that needs to be transferred to the central server and to provide a better data management scheme to adapt the heterogeneity. Another set of researchers integrated FL with imitation learning (IL) and proposed a framework for improving the performance and the learning capability of a robot\cite{liu2020federated}. The framework was designed for cloud based robotic systems which gathered sensor data from heterogeneous resources. \textcolor{black}{Yin \emph{et al.}} \cite{yin2020fdc} proposed a framework based on FL for providing a secured collaboration of data. The framework was named as FDC and enabled data collaboration among multiple parties distributed across different geographical locations.

\textcolor{black}{Lu \emph{et al.}} \cite{lu2019blockchain} designed a methodology by integrating the blockchain and FL for enhancing the computational efficiency and to secure the sharing of data among the distributed clients as well as the server. This work provided an IIoT test bed for analyzing the performance of their framework. FL is used for solving the data sharing problem in typical IIoT environment. The presence of a centralized server gave rise to so many data leakage problems. Most of the clients hesitated to share their private raw data over the network to a centralized location. Hence, a methodology using FL was designed in which the model or the structure of the data was shared to all clients in the IIoT Network instead of leaking or revealing the raw data. 

\textcolor{black}{Zhu \emph{et al.}} \cite{zhu2019multi} proposed a methodology to optimize their neural network model with the usage of evolutionary algorithms. This work used multi-objective functions for reducing the computational cost. FL is used for the purpose of solving the scalability issues in the DL models and also to increase the learning efficiency and performance of the global model using the data distributed across the clients.

A methodology called two-phase multi-party computation (MPC) was proposed by Renuga \emph{et al.} which is an aggregation model using FL concept \cite{kanagavelu2020two}. The model provided privacy to huge sensitive data gathered from various companies. This followed a two phase methodology where initially a small committee was elected and later MPC based aggregation service was provided via the elected committee. The model was integrated along with a smart manufacturing IoT environment and the results showed high accuracy with reduced communication cost and computation time. 

FL is now integrated with DL models to provide intelligence to industrial applications. A scheme by name Efficient and Privacy Enhanced FL (PEFL) is proposed in \cite{hao2019efficient}. This scheme is best suited for industrial sectors dealing with sensitive data like healthcare, auto-pilot, auto-driving, industrial robots, etc. where decision making cannot be compromised due to hesitation in sharing data. The major advantage of this scheme is that the performance is not reduced even if there is a collision due to multiple entries.  

A cross domain sharing based scheduling scheme\cite{zhang2019edge} was proposed for providing edge services with higher intelligence. The scheme integrated the block-chain and edge intelligence mechanisms to provide an intelligent IIoT framework. The results showed an improvement in service cost of the edge devices and their efficiency in terms of service. The proposed scheme is suitable for IIoT networks beyond 5G. This work also designed a transaction approval scheme for reaching the edge resources which are distributed all over.

An optimization technique using particle swarm optimization (PSO) and FL is proposed in \cite{qolomany2020particle} focusing on the hyper-parameter tuning for the DL models that are available locally in the smart city applications. PSO is used for setting the hyper-parameters in case of the DL model LSTM. The results showed a reduction in number of rounds required for identifying the parameters by 2\% to 4\%.

To summarize, there are various methods and building blocks that can be used for customizing the traditional ML and FL algorithms to suit the IIoT applications. Though there are too many architectures available for FL, there is no specific architecture that seems to be centered around specific industrial sector. The algorithms and methodologies discussed in the section can work for any type of industry once modified a bit keeping in mind various aspects such as efficiency, device set up, end users, end servers, structure, autonomy. etc. 

\subsection{Federated Learning for Resource Management}

The recent industrial revolution uses a wide variety of technologies like collaborative robotics, cloud and edge computing, cognitive computing, CPS, ML, etc.\cite{tao2019digital} These collaborative technologies have led to high level of industrial innovations and applications. Integration of Industrial applications with IoT and ML leads to various advantages like optimized production costs, increased productivity, reduction of error, improvising the automation process, higher quality, and so on. Hence ML plays a major role in any smart industrial application. But, the traditional ML aspects could not be used in real-time Industry 4.0 applications as data \textcolor{black}{cannot} be shared for training in a centralized ML server\cite{lim2020federated} as there is a risk of data leakage\cite{learning2020step}. Hence most of the \textcolor{black}{IIoT} uses FL which is \textcolor{black}{an efficient resource management} technique for learning, adding new data and updating the aggregation server with the model updates. This section discusses about the various contribution of researchers in the field of FL for optimizing and managing resources over real-time processing in Industry 4.0 applications.

In \cite{chen2020joint}, Chen \emph{et al.} analysed the performance of FL with the help of packet error rate over wireless networks. A closed-form expression was derived for calculating the convergence rate of FL. This was utilized for computing the optimal transmission power for allocating a resource block and selecting a user. This work also minimized the loss function for the above said resource allocation and user selection. 
This work was analyzed using a single base station (BS) scenario, and thus the blooming researchers can concentrate on analysing the performance of FL for multiple BS scenarios.
In \cite{abad2020hierarchical}, \textcolor{black}{Abad \emph{et al.}} proposed hierarchical FL for optimizing the usage and allocation of resource in wireless applications. This work constructed a heterogeneous cellular network which consisted a macro base station (MBS) and few small cell base stations (SBS). In the SBS, sub-global model is trained using the traditional FL concept for specific iterations and this process is carried out in all SBS. The trained sub-global models in all SBS are then transferred to the MBS for aggregation of the global model. The global model updates are finally transferred to the SBS from where the updated parameters of the global model are broadcasted to the end devices. This technique is well suited for reusing the wireless resources. Though this seems an attractive technique, the scheme follows a centralized MBS which might crash due to physical damage or attack and hence there is a chance of losing the updated global parameters.

\textcolor{black}{Huang \emph{et al.}} \cite{huang2019deep} proposed a novel approach named Deep-Q for the purpose of enabling resource allocation and task offloading for networks in edge computing. This work focused on offloading multiple tasks at the same time to the edge servers. The proposed technique helped in minimizing the cost due to delay, computation and energy consumption. 
In \cite{khan2020resource}, \textcolor{black}{Khan \emph{et al.}} proposed a novel dispersed FL to be employed in smart industries. This work focused on optimizing the resource in cognitive IoT based smart industrial applications. This work proposed a model which used distributed and hierarchical FL for offering optimized resource communication and robustness. This work also formulated an integer linear programming problem for minimizing the overall cost of FL. Numerical results were provided to compare the proposed approach with two techniques namely random resource allocation and random association.

\textcolor{black}{Hiessl \emph{et al.}} \cite{hiessl2020industrial} tried to adapt the available traditional FL for industrial environments and named it as Industrial FL (IFL). This work introduced the various requirements and designs required for IFL. IFL trains and evaluates the ML models. This work also discussed the needs for IFL and also provided a collection of structured workflows for the requirements in the architecture of IFL. To get adapted to the diverse operating conditions of the industrial environment, the FL clients are not allowed to exchange the ML model parameters with the FL participants in the IFL architecture. A discussion on future such as usage of FL open source frameworks, asynchronous and decentralized FL, and negative knowledge transfer were also provided.

In \cite{lu2020low}, \textcolor{black}{Lu \emph{et al.}} introduced the usage of blockchain based FL in the digital twin wireless networks (DTWN). The methodology helps in improving the reliability, security and data privacy by which the learning accuracy is balanced using the bandwidth allocation process. The designed DTWN model uses the digital twins concept for mitigating the unreliable communication among the edge servers and users. The limited resources available in the network are allocated using the optimization problem that is formulated between the BS and the digital twins. The major advantage of the scheme is good learning convergence and latency.

\textcolor{black}{Messaoud \emph{et al.}} \cite{messaoud2020deep} focused on developing a FL based methodology to support the resource allocation strategy in multi-industrial IoT. This work proposed a novel deep federated reinforcement learning (DFRL) scheme for providing dynamic resource allocation and network management for the IIoT networks. The scheme provided IIoT slice based resource allocation with the help of a novel proposal Deep Federated Q-Learning (DFQL). This work also proposed two other approached namely Multi-Agent deep Q-learning (MAQL) that dynamically slices the transmission power and spreading factor for maximizing the QoS requirements namely throughput and delay. The DFL is also utilized for learning the multi agent model to enable decision making on the IIoT virtual network slices. 

Edge intelligence plays a significant role in IIoT since it provides smart services via cloud with low latency rate and lesser cost. Since IIoT is a heterogeneous network with diverse resources, there is a need to integrate all these together to improve the computing capability. \textcolor{black}{Zhang \emph{et al.}} \cite{zhang2019edge} proposed a scheme for cross-domain sharing of resources by using an edge scheduling scheme based on DFL. This scheme is designed based on a credit-differentiated edge transaction approval mechanism which enhances the flexibility and security in the service provided by the edge with a minimal cost.

\begin{table*}[!ht]
\centering
\caption{Summary of Recent Research Contributions on FL in IIoT for Data Storage-Data Management-Resource Management.}
\label{tab:summary}
\resizebox{\textwidth}{!}{%
\begin{tabular}{|l|l|p{3.5cm}|p{3.5cm}|p{5cm}|p{3.5cm}|}
\hline
Ref. &
  Methodology &
  Research Focus &
  Techniques Used &
  Contributions &
  Proof of Concept \\ \hline
{\cite{fu2018secure,borylo2016energy,kaur2018edge}} &
  \multirow{12}{*}{\makecell{Data Storage}} &
  
  $\circ$ Secure data storage 
  
  $\circ$ Dynamic data gathering 
  
  $\circ$ Reducing congestion 
  
  $\circ$ Reducing energy consumption &
  
  $\circ$ Middleware based solution 
  
  $\circ$ SDN 
  
  $\circ$ Interplay of fog and cloud computing 
  
  $\circ$ latencyAware policy &
 
  $\circ$ Retrieval feature tree is designed 
  
  $\circ$ Reduction in storage space 
  
  $\circ$ Carbon footprint of data centre is reduced 
  
  $\circ$ Multi-objective evolutionary algorithm is proposed &
  $\circ$ Gas density in factory 
  
  $\circ$ Customer services and maintenance \\ \cline{1-1} \cline{3-6} 
{\cite{singh2018fuzzy,saqlain2019framework}} &
   &
  $\circ$ Effective data storage 
  
  $\circ$ Acquiring huge data 
  
  $\circ$ Provide intelligent management &
  $\circ$ SOA based solution 
  
  $\circ$ Distributed database server 
  
  $\circ$ Data aggregation model &
  $\circ$ FFBF service architecture is proposed 
  
  $\circ$ An IDMS framework is proposed &
  $\circ$ Manufacturing sector 
  
  $\circ$ Healthcare 
  
  $\circ$ Wearable devices \\ \cline{1-1} \cline{3-6} {\cite{anton2019highly}} & &
  $\circ$ Collection of data from various  sources 
  
  $\circ$ Aggregation of data
  
  $\circ$ Analysis of data &
  $\circ$ Event Centric Model &
  $\circ$ A data aggregation model is proposed
  
  $\circ$ Correlation model is proposed &
    $\circ$ Maintenance Sector 
  
  $\circ$ Interconnected Production Lines  \\ \cline{1-1} \cline{3-6}
  
{\cite{wan2016software,shu2016cloud,liu2018blockchain}} &
   &
  $\circ$ To handle device failure 
  
  $\circ$ To provide interface for exchange of   information and data &
  $\circ$ Architecture based solution 
  
  $\circ$ SDN 
  
  $\circ$ CPS 
  
  $\circ$ DRL &
  $\circ$ Blockchain based DRL is proposed 
  
  $\circ$ CCPSA is proposed 
  
  $\circ$ Equipment utilization rate is increased &
  $\circ$ Complex industrial applications 
  
  $\circ$ Automated guided vehicles 
  
  $\circ$ Industrial robots \\ \hline
{\cite{wang2019edge}} &
  \multirow{29}{*}{\makecell{Data\\ Management}} &
  $\circ$ To reduce the amount of data uploaded 
  
  $\circ$ To provide intelligence to mobile edge nodes &
  
  $\circ$ DRL for optimization 
  
  $\circ$ FL for providing intelligence &
  $\circ$ An “In-Edge AI” framework for improving deployment 
  
  $\circ$ Deep Q-Learning for improving   computation in edges &
  Xender’s traces \\ \cline{1-1} \cline{3-6} 
{\cite{liu2020federated}} &
   &
  $\circ$ To provide knowledge fusion mechanism 
  
  $\circ$ To provide a model for transfer learning &
  FL along with imitation learning &
  $\circ$ Federated IL (FIL) is proposed 
  
  $\circ$ A transfer learning approach using FIL is proposed &
  Self-driving cars \\ \cline{1-1} \cline{3-6} 
{\cite{yin2020fdc}} &
   &
  $\circ$ To provide higher security while multiple parties try to collaborate &
  DL and FL along with blockchain &
  $\circ$ A data collaboration framework is proposed using FDL 
  
  $\circ$ A blockchain based secure mechanism for computation is proposed &
  $\circ$ A private blockchain based on Libra protocol 
  
  $\circ$ Raw data collected from wearable device \\ \cline{1-1} \cline{3-6} 
{\cite{lu2019blockchain}} &
   &
  $\circ$ To share data across multiple parties without any leakage &
  FL and blockchain &
  $\circ$ Differential private FL is proposed &
  $\circ$ Reuters data set  
  
  $\circ$ 20 newsgroups data \\ \cline{1-1} \cline{3-6} 
{\cite{zhu2019multi}} &
   &
  $\circ$ To optimize the structure of the model 
  
  $\circ$ To minimize the computational cost 
  
  $\circ$ To minimize the global model test errors &
  FL and multi-objective evolutionary algorithm &
  $\circ$ Federated averaging algorithm is proposed 
  
  $\circ$ A modified version of SET algorithm is  proposed &
  Independent IID and non-IID data set \\ \cline{1-1} \cline{3-6} 
{\cite{kanagavelu2020two}} &
   &
  To provide better privacy in shared models &
  FL and MPC &
  $\circ$ Two-phase MPC enabled FL is proposed 
  
  $\circ$ Committee election algorithm is proposed &
  $\circ$ Manufacturing unit 
  
  $\circ$ Fault detection in electrical machines \\ \cline{1-1} \cline{3-6} 
{\cite{hao2019efficient}} &
   &
  To prevent exploitation of private data   in sensitive applications &
  FL along with Augmented Learning &
  $\circ$ Privacy enhanced FL is proposed for Industrial AI 
  
  $\circ$ Aggregation decryption is utilized 
  
  $\circ$ Model aggregation algorithm is proposed &
  $\circ$ MNIST data set 
  
  $\circ$ Healthcare and autopilot \\ \cline{1-1} \cline{3-6} 
{\cite{zhang2019edge}} &
   &
  $\circ$ To provide intelligence to the edge network 
  
  $\circ$ To prevent malicious attacks &
  Ubiquitous communication, blockchain along with DRL &
  $\circ$ Credit Differentiated transaction approval scheme is proposed 
  
  $\circ$ DRL for optimized resource scheduling &
  Five BS each having computing server and caching server \\ \cline{1-1} \cline{3-6} 
{\cite{qolomany2020particle}} &
   &
  To optimize hyper-parameters in ML model &
  PSO along with FL &
  $\circ$ PSO based parameter setting framework is proposed 
  
  $\circ$ The number of hidden layers and neurons are optimized &
  Smart city traffic with City Pulse EU FP7 data set \\ \hline
\cite{chen2020joint,abad2020hierarchical} &
  \multirow{15}{*}{\makecell{Resource\\ Management}} &
  $\circ$ To optimize transmission power 
  
  $\circ$ To reuse wireless resources &
  Hierarchical FL &
  $\circ$ Closed form expression for computing convergence rate 
  
  $\circ$ Optimal transmission power 
  
  $\circ$ Minimized loss function &
  $\circ$ Single BS scenario 
  
  $\circ$ Heterogeneous cellular network consisting of MBS and SBS \\ \cline{1-1} \cline{3-6} 
{\cite{huang2019deep}} &
   &
  To enable task offloading in edges &
  DL based FL &
  $\circ$ Minimize cost due to delay, computation  and energy consumption &
  Heterogeneous networks in edge computing \\ \cline{1-1} \cline{3-6} 
{\cite{khan2020resource}} &
   &
  To optimize resource in cognitive IoT &
  Dispersed FL &
  Integer linear programming to minimize overall cost &
  Smart industry \\ \cline{1-1} \cline{3-6} 

{\cite{hiessl2020industrial}} &
   &
  $\circ$ To evaluate ML model
  
  $\circ$ To provide structured workflows&
  IFL &
  $\circ$ Model for adapting to diverse operating conditions
  
  $\circ$ Proposed IFL&
  $\circ$ Synchronous framework
  
  $\circ$ Centralized FL\\ \cline{1-1} \cline{3-6}   
  
{\cite{lu2020low}} &
   &
  To mitigate unreliable communication &
  Blockchain based FL &
  Learning accuracy is balanced &
  DTWN \\ \cline{1-1} \cline{3-6} 
{\cite{messaoud2020deep}} &
   &
  To provide dynamic resource allocation &
  DFRL &
  MAQL for dynamic slicing is proposed &
  Multi-industrial IoT \\ \cline{1-1} \cline{3-6}
 {\cite{zhang2020federated}} &
   &
  To perform image classification in exploration domain &
  Distributed FL &
  $\circ$ GFC is deployed
  
  $\circ$ Communication cost is reduced
  
  $\circ$ Weighted zero forcing transmit precoding &
  UAVs \\ \hline
\end{tabular}%
}
\end{table*}

Unmanned aerial vehicles (UAVs) play a promising role in IIoT applications, but the characteristic of UAVs is that they have a limited payload and lesser flight time \cite{pham2020sum}. These UAVs have to be empowered with complex as well as high level of tasks so that they can be utilized in smart IIoT applications. In \cite{zhang2020federated}, \textcolor{black}{Zhang \emph{et al.}} investigated the UAVs to perform image classification tasks and proposed FL based multi-UAV systems in the exploration domain. The image classification task is completed using the distributed FL. Multiple UAVs in the application area are coordinated with the use of ground fusion centre (GFC). Results show that the accuracy of image classification by UAV is high and communication cost is reduced relatively. The global model is updated at the GFC using the weighted Zero Forcing transmit precoding at every UAV.

\begin{figure*}[t]
	\centering
	\includegraphics[width=1.00\linewidth]{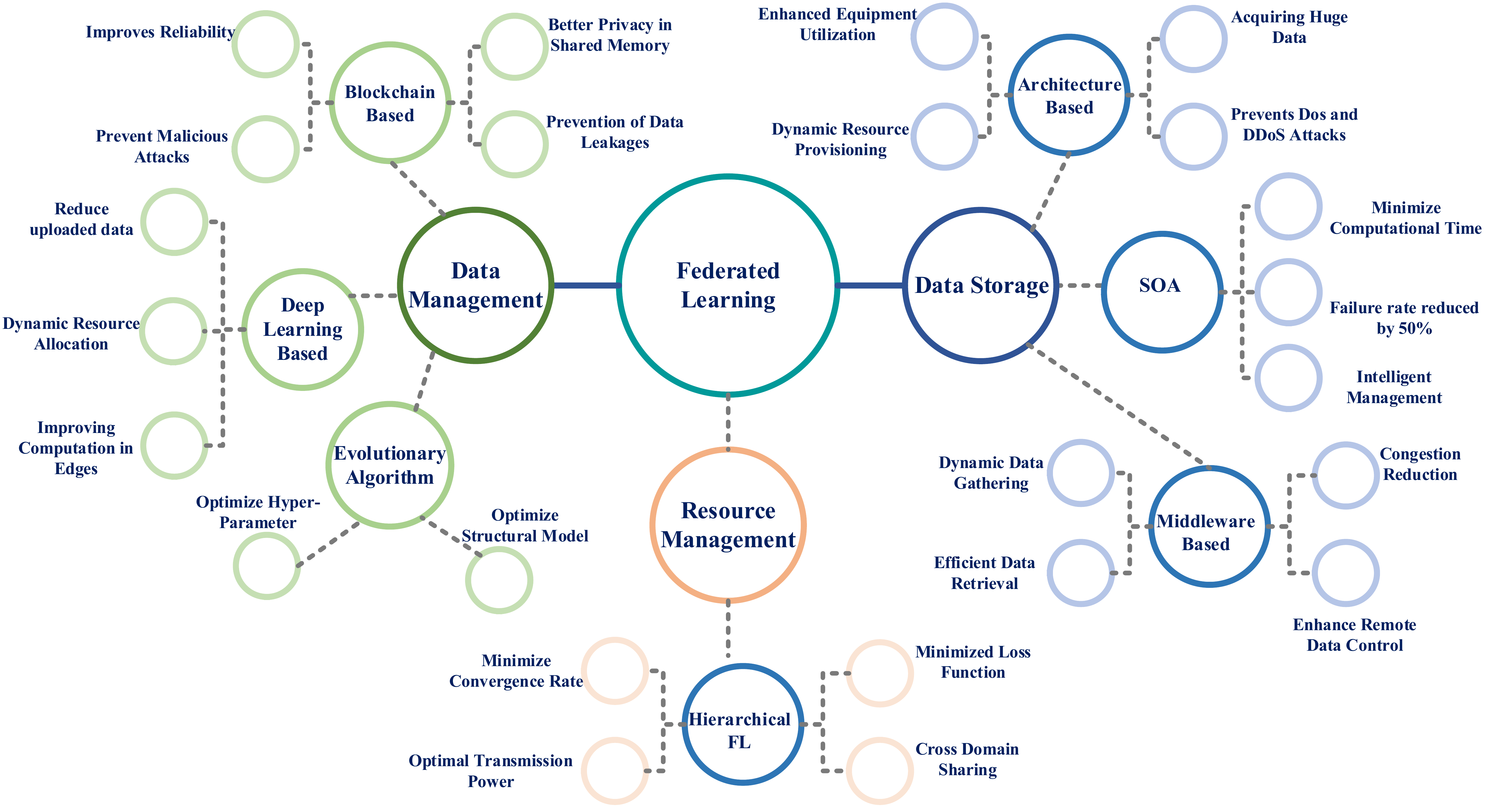}
	\caption{\textcolor{black}{Illustrative Usage of FL in IIoT for data/resource management.}}
	\label{Data Management}
\end{figure*}

To summarize, resource optimization deals with optimizing the communication as well as computation resources required for enabling the FL systems to locally compute, communicate and update the global parameters. When these are optimized, the learning cost of FL model is significantly reduced. Apart from these the energy consumed by the end devices and local learning rate are dependent on the capability of FL systems' CPU. Various research contributions discussed in the section gives solutions for reducing the computation cost and improvises the resource allocation process. 

\textcolor{black}{It is expected that IIoT would transform the way human beings live, work or play. The future of IIoT varies from basic automation to factory automation, high level of connectivity to wearable devices and smart applications. IIoT is ready to set our lives with large number of sensible things around us which would be networked with each other to avoid any accidents or abnormalities. In the near future, there will be a variety of applications which would be a part of our daily activities. There will be huge volume of data acquired by networks, upload and download of data via communication channel would be high. To handle these data, various solutions are proposed by recent researchers with respect to data storage, data management and resource management and is summarised in the Table \ref{tab:summary}. The usage of FL for data/resource management and data storage in IIoT is illustrated in Fig.~\ref{Data Management}.}

\section{Applications }

IoT has been playing significant role in our day to day modern life with applications almost in every vertical such as smart automobile industry, smart healthcare, defence, smart cities, smart grid etc \cite{atzori2010internet}. It provides high end smart services and products which can transform traditional technology into an fully automated technology. Typically, millions of smart sensors are producing bundles of data that needs to be managed specifically privacy as per new GDPR rules. As a consequence, task-specific ML models need to be developed that can provide smart services and products with no compromise on privacy.

\label{sec:Applications}
\subsection{Federated Learning enabled IoT in Automotive Industry}
FL is a ML approach that trains an algorithm in a decentralized manner such that models are updated with the help of local data without sharing this data over cloud servers keeping the privacy of the data intact \cite{mcmahan2017federated}. FL along with IoT has become the game-changer for almost the last two decades in every modern field such as smart automobile industry, smart healthcare, defence, smart cities, smart grid etc \cite{atzori2010internet}. These collaborative techniques along with IoT have transformed the way every individual interacts with his surroundings by providing smart services which convert traditional technology into a fully automated technology \cite{pokhrel2020towards}. Considering automobile industry, the consumer end has been benefiting the most as modern vehicular systems have been transformed into an automated device which enables all the smart vehicles to be interconnected and also occupants in these vehicles to become connected to each other. Vehicles that are connected to IoT provides constant and simultaneous access to information to chauffeur and occupants while moving. However, the number of vehicles connected to each other over IoT keeps on increasing \cite{james2014internet} which thrives for smart car emerging technologies with new requirements such as secure, robust and roadside infrastructures etc, which  transforms the original concept of vehicular ad-hoc networks (VANETs) \cite{zeadally2012vehicular}, \cite{isaac2010security}, \cite{guerrero2013vehicular} into a new concept called the FL enabled Internet of Vehicles (IoV) \cite{du2020federated}. Manufacturing of such smart vehicles in automotive industry must be capable of employing IIoT along with ML techniques paving the way for smart industry experience enabling efficient and sustainable production\cite{zolanvari2019machine}. Computers, people, and machines coming together to execute industrial operations by capitalizing advanced data driven algorithms  for groundbreaking business results is termed as IIoT. 
IIoT provides a way to better understand and gain insight into operations and assets of the company with the help of data associated with the various machine sensors, middleware, software, and  back-end  cloud  compute and  memory  systems. Therefore, it provides a way to transform business operations and processes by using advanced analytics to query large data sets by using the results obtained as feedback. These gains results in optimized efficiency of productions, and thereby increasing profits. This connectivity allows for data collection, exchange, and analysis, potentially facilitating improvements in productivity and efficiency. IIoT applications usually require relatively small throughput per node, and capacity is not a major issue. Instead, a large number of devices need to be connected to the Internet at low cost, which have limited hardware capabilities and have small batteries thus requires an energy efficient, cost effective, reliability, and privacy solutions. The IIoT will eventually have a greater impact on economy as it will innovate infrastructures to such larger extents that it will revolutionize the traditional industry. With a strong focus on M2M communication, big data, and ML, the IIoT enables industries and enterprises to have better efficiency and reliability in their operations. As big and high  volumes of data is involved in IIoT devices, it is impossible to collect data over cloud centralized server \cite{niknam2019federated}. As high number of devices and machines are connected, classical cloud-based learning approaches are impractical and inadequate. But with FML enabled IIoT, privacy of the data is taken into consideration which further enhances the performance of IIoT devices and also accelerates the learning processes of the data which technique you are talking or building the base. 

There are a number of techniques available to access and train data. One such approach is to access user data which is scattered and collecting it from cloud for modeling and then transfer this trained model to client devices for providing automation of specific tasks. But the cost associated with this type of approach is huge and also latency will be increased which is not desirable\cite{zhou2019edge}. It also involves some data privacy issues as sensitive data is shared on cloud which can be leaked. The IIoT devices are also becoming the main targets of malicious attacks leading to disastrous consequences. These attacks misleads the global models to predict undesirable outputs \cite{xu2018enabling}. Various counter methods have been proposed such as ensemble diversity \cite{pang2019improving}, Purifying Variational AutoEncoder (PuVAE) \cite{hwang2019puvae} and other training methods such as adversarial training, but none of them proved to be satisfactory as they focus only on a particular type of attacks. Alternative to these methods is to train the IIoT data locally in isolation to other devices but with this approach, resources required will be very high. On the other hand, with limited computation ability and local memory and insufficient data samples can result in further degrading the model. So to solve the issue of availability of IIoT device data in isolation and maintaining privacy of this data, a more advanced ML approach is developed which is termed as \textit{Federated ML} \cite{yang2019federated}. The concept of FL was first introduced by google in 2016 \cite{konevcny2016federated} which aimed to access user data distributed across various IIoT devices and build ML models exploiting this data without its leakage \cite{bonawitz2017practical}. Privacy is one of the basic attributes of FL which requires security models and analysis to provide meaningful privacy guarantees \cite{geyer2017differentially}. FL enabled smart automobile industry has various automation systems, IIoT devices for storing corresponding data, control systems and various operating machines which are connected to each other through the internet. FML models are deployed to edge devices which updates the model based on demands and also check for any anomalies in production\cite{husakovic2018robust}.

\begin{figure*}[ht]
\centering
  \includegraphics[width=1.0\linewidth]{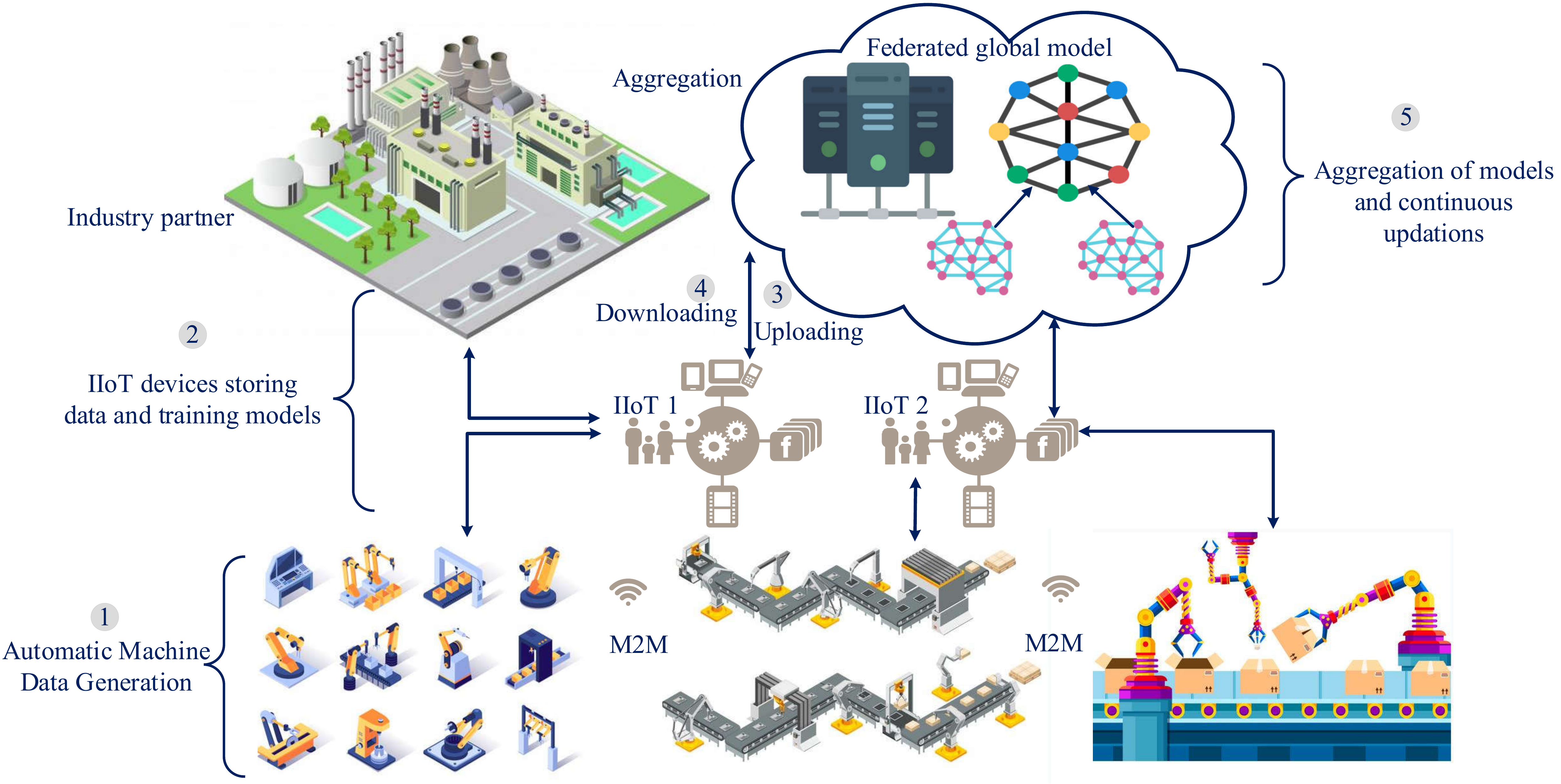}
  \caption{A FL-enabled smart automobile industry manufacturing smart cars. FL models are being trained by edge IIoT devices which trains data generated by automatic machines and the data fetched from industry partners. These models are regularly updated and sent to global cloud for aggregation and generating a new global model which is reliable and efficient.}
  \label{fig:Fig 1}
\end{figure*}

An FL based IIoT model is depicted in Fig.  ~\ref{fig:Fig 1} and steps followed are mentioned below.
\begin{enumerate}
    \item \textit{Selection of client nodes}: In a smart industry, various clients are automation systems and operating machines having its own processor and is connected to other machines having M2M communication, the IIoT subsystems collecting data from data generating machines for learning tasks and training of models\cite{mcmahan2017communication}. These client edge nodes are selected by central cloud server that is involved in model training. This big data associated with other smart industries of some other company can also be considered while training the global model. Based on \textit{client selection}, an initial central model is developed by evaluating local raw data of edge clients.
    \item \textit{Model distribution}: This initial model is then distributed to clients for the purpose of updating and training this model based on clients local raw data. 
    \item \textit{Uploading locally trained models}: These initial models are no longer initial as they are trained by the clients using their local raw data. The updated model is then uploaded to centralized server for federated averaging.
    \item \textit{Aggregation}: Having all the updated models by local clients, central task executioner averages all these models to develop a new version of Global model which is better in terms of efficiency and accuracy than the initial global model. For averaging FedAvg \cite{mcmahan2017communication} is the best algorithm to solve this averaging problem yielding a well trained Global model. 
    \item \textit{Model Testing and update}: This aggregated global model is then tested by central task executioner with the data that belongs to other clients that have not participated in this training. Analyzing the testing results, some parameters are tuned to repeat the training process if needed, else the model is updated accordingly and distributed to all the clients.
    
\end{enumerate}

The above process is repeated until a global model is achieved which solves the purpose with considerable accuracy \cite{anh2019efficient}. Based on this, Bonawitz et al. \cite{bonawitz2017practical} recommend that all local updates are aggregated securely and globally by using their weighted average. In this way, efficiency in production can be achieved with low cost.

In smart automobile factory, automated machines and control systems drives the raw material used for production by communicating with each other cooperatively. Smart manufacturing enables factory managers to automatically collect and analyze data to make more informed decisions and optimize production in reduced costs. All the decisions depends upon harnessing of data and based on FL models, data will decide what to do and when to do a particular task.
\subsubsection{Requirements}
Following are the requirements for FL enabled IIoT in automotive industry for continuous monitoring and adaptation of system. 
\begin{itemize}
    \item \textit{Data management}: First and foremost requirement is the management and harnessing of big data. To support collaboration of all the FL clients (FLCs), it is required to publish metadata describing the organization and its corresponding edge IIoT device. Based on this data, FLCs can provide information and data for collaborating with other selected FLCs. As private raw data is not shared in FL, it enables privacy taking into consideration the privacy policies of industry partners. Industrial FLCs target to improve ML models using IIoT device data by updating them regularly with its local data.
    
    \item \textit{M2M communications}: These are required in automated machines so that they can send and receive data from other machines and can work collaboratively resulting in smart manufacturing of vehicles. Machines communicate with each other through various protocols in production and manufacturing of smart vehicles. In case of M2M communications, efficient routing protocols are required in dense and mobile environment \cite{yang2020v2v}, \cite{wu2018computational}. The driving force behind the FL-based IIoT is that smart machines with M2M communications works better than humans with greater accuracies by consistently capturing and communicating data with each other.
    
    \item  \textit{Quality of Information (QoI)}: Quality of information means the information set that is received by global model through aggregations and updations by local models with their local data sets. As FLCs trains and evaluates the models based on its local data set, aggregated global models have a diverse set of data with certain quality of information. Furthermore, as different clients are operating in the industry with less or almost no human interventions \cite{jennings1994archon}, different approaches to collect data can influence QoI of generated data sets in edge IIoTs such as free-of-error datasets, relevancy QoI, etc \cite{lee2002aimq}. Storing QoI indicators along with existing industry metadata of participating organizations can further enhance the ability to build and update appropriate FL models.

    \item \textit{Scheduling Algorithms}: These algorithms are required for computations and task offloading as FL plans can cause heavy loads on edge IIoT devices as training requires large amount of data sets and this problem was identified in \cite{bonawitz2019towards} and the need for scheduling algorithms. These algorithms take care of any repetitions being done on older data sets if any new data is to be processed and model should be trained on the basis of new data. The scheduling device will give priority to this new data set and avoid any congestion. 
\end{itemize}

\subsubsection{Advantages}

The FL enabled IIoT is proved beneficial in industry in the following ways:
\begin{itemize}
    \item A FL approach which is dynamic and decentralized approach, can be \textit{efficient} in terms of design and allocation of resources by collaborative and simultaneous learning with constant access of information from automated machines and control systems releasing its data to IIoT edge devices. 
    \item FL enabled IoT is said to take care of privacy of the data which the client has to constantly share with the task executioner for generating and updating global model thus mitigating the privacy issues by not providing access to the raw training data.  Privacy-protection techniques can be applied to FL such as personalization techniques \cite{kulkarni2020survey}, FL enabled semi-supervised learning algorithms \cite{jin2020survey}, threat models \cite{lyu2020threats}, mobile edge networks \cite{lim2020federated}. By using large amount of data for devices with privacy protection methods, FL can enable some impossible new applications for traditional learning methods.
    \item FL approach enables smart manufacturing to take actions on the basis of global model and local model which results in reduced delay as compared to traditional techniques. The FL can be incorporated with other techniques such as Blockchain and edge computing to promote various real-time systems for the IoT in Industries.
\end{itemize}

\subsubsection{Applications}
FL enabled IoV is an autonomous system which updates and trains global model in isolation, thus maintaining the privacy of the user data. Existing IoT vehicular systems are based on centralization of data in which each vehicle uploads its data to a central cloud which puts a limit on amount on data being shared as there is a risk of data leakage. But, FL mitigates this issue and intelligent autonomous vehicular system could be developed which works on collaboration among vehicles. Thus it has many applications such as intelligent traffic signal control, precise navigation systems, collaborative autonomous driving, collision avoidance systems, vehicle platooning and many more.

\textit{Vehicle platooning system} is very important application of ML based IoT systems. For building smart cities, autonomous vehicle platooning will play an important role in maintaining the road safety and mitigates the problem of traffic congestion. FL enabled IoT along with vehicle-to-vehicle (V2V) communication will collect information from environment and trains global model without any risk of data leakage, thus maintaining the speed requirements and the V2V distance and provide stability to vehicle's control system. Modern Vehicles are equipped with multiple sensors, and accessing data of multiple vehicles along with FL, improved platooning systems can be achieved \cite{zeng2019joint}. GPS, speed of vehicle, V2V distance data can be accessed for predicting road safety, enabling a better positioning systems and intelligent road systems \cite{ferdowsi2019deep}. The camera of the vehicle and data associated with it can be accessed for surveillance, any accidents etc. However, due to large data size and privacy concerns, it is challenging to upload the data on central cloud server. Therefore, FL could be a possible way to conduct distributed perception by processing the data at the end vehicles and then integrating the results to make the final decision.

\begin{figure*}[ht]
\centering
  \includegraphics[width=1.00\linewidth]{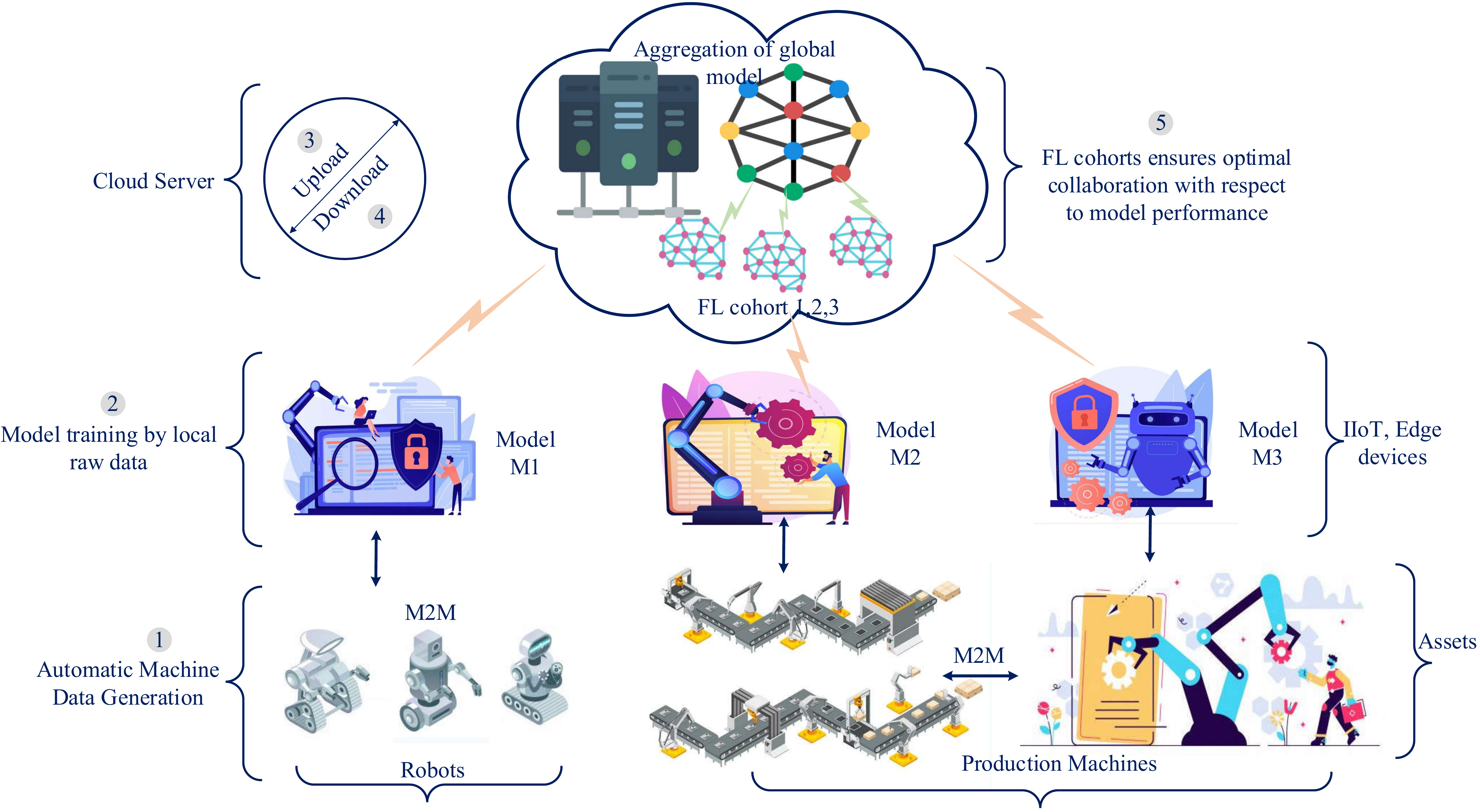}
  \caption{An FL-Enabled IoT healthcare system.}
  \label{fig:Fig 2}
\end{figure*}

\subsection{Smart Health Care Industry}

Healthcare industry is another area that we expect to benefit from the rising ML technique of FL. As the population is growing at alarming rate, there can be seen an alarming pressure on hospital staff and quality of service being delivered to patients by staff and medical practitioners. Thus, new efficient and reliable solutions are required from science and technology\cite{challoner2019intelligent}, \cite{omanovic2015future}.Because of fourth industrial revolution, innovation has been achieved in manufacturing process with increased efficiency in the supply chain \cite{zhang2019edge}. In the medical field, smart Industry with IIoT and ML techniques has played vital role providing wide range of applications in creating customised implants, tools and medical equipment. Smart healthcare industry improves the healthcare infrastructures and biomedical systems keeping in view the safety of patients, their vitals and real time monitoring. Such automated systems manufactured by smart medical industry make smart decisions through real-time communication and  cooperation  with  humans, machines, sensors, and so forth. \cite{wang2016towards}. Medical Industry 4.0 manufactures smart devices which provides quality care to patients while addressing the issue of shortage of medical staff and beds especially in populated hospitals. IIoT has been widely identified as a potential solution to alleviate the pressures on healthcare systems \cite{gope2015bsn} by assisting manufacturing and supply chain of medical devices. Remote-healthcare systems can monitor patients at home without any need of them visiting infirmary or hospitals. In essence, it can improve access to medical resources, while reducing the pressure on the medical system, and it can allow people to better control their health. Health systems have improved a lot from the earlier times, thus increasing average life expectancy of the individuals. As a result, there is increased population of elderly people living with chronic diseases, heart diseases, diabetes etc, thus placing huge demands of healthcare services. In order to address this issue, FL enabled IIoT systems are developed which can learn patients medical data such as day-to-day health, disease symptoms, medical reports and vitals and this data is shared with medical practitioners to subscribe them appropriate medication at their residence itself. There are a lot of IIoT based healthcare systems that focus on particular diseases and in-general monitoring of patients health. A system for generating rehabilitation plan based on patients symptoms has been developed \cite{gradim2020iot} in which the individual’s condition is compared with the symptoms of past patients from a database and studied their ailments and provided optimal treatments. IIoT provides anytime, anywhere and anything can be connected with each other with focus on M2M communication \cite{atzori2010internet}. IIoT and IFL enable ML models on local raw data, stored in edge devices, to be updated to global model without actually sharing the local data which provides security to manufacturing process and IIoT improves productivity \cite{shokri2015privacy}. In this scheme, cloud server receives different parameters from local updated models trained on IIoT decentralized \cite{pokhrelfederated} edge devices and aggregates these local models to develop a global model \cite{mcmahan2017communication}, whose parameters are again sent to edge devices for training based on raw data. This process is repeated until the global model achieves a well defined condition and accuracy. But to successfully caters the challenges faced in 
industry, FML approach needs to be adapted. 

Thus FL along with IIoT is employed which keeps all the sensitive data of automated systems and smart operating machines private in local edge devices. The machines are digitally represented as \textit{assets} \cite{hiessl2020industrial} that are responsible for generation of data that is trained and evaluated by edge devices.

The FL enabled smart healthcare industry model is depicted in Fig.~\ref{fig:Fig 2} and following steps are followed:
\begin{enumerate}
    \item \textit{Data Selection}: Central cloud server needs an initial set of data to develop an initial global model. This data set includes the data that is generated by \textit{assets} which depicts automated machines. This data is accessed by edge devices which trains models based on ML. ML models depicts any anomalies present in the manufacturing process of the smart medical devices. Due to some potential differences in asset data, model updates may lead to negative knowledge transfer, thereby reducing model performance \cite{tuli2020next}.

    \item \textit{Model distribution and updation}: This initial model prepared by cloud is distributed to IIoT edge devices which can update and train this model based on their own local data. Due to the existence of well-trained robots, the ML model are constantly updated and the information is shared with other federated robots. These updates are integrated asynchronously into the global model to ultimately enhance the navigation skills of all robots involved. In this way IFL can continuously updates and reevaluate data similarity which is key requirement to ensure high model quality.

    \item \textit{Aggregation}: Algorithms such as FedAvg \cite{mcmahan2017communication} is used to solve this averaging problem yielding a well trained Global model. It is decentralization approach that work with non independent identical distribution (non-IID) data which requires sharing of the same model by all the clients.
    
    \item \textit{Federated Learning Cohorts}:  FL cohorts enables clients or \textit{assets} share their updates to those belonging to subset of FL clients, who have submitted FL tasks that belongs to the same FL cohort. These updates are required to improvise accuracy of their local models. In medical device manufacturing industries, some assets requires similar conditions for their accurate working e.g., automated production machines placed into environment with similar temperature, noise and other features considered in the model prediction. Thus the IFL system needs to build FL cohorts.

    \item \textit{Scheduling}: A service scheduling machine and a virtual manager (VM) is also required at the middle of IoT subsystems and cloud. VM is responsible for collecting and storing sensor data of virtual sensors from automatic machines and robots. Service scheduling machine is responsible for resource management and scheduling of several tasks with high inter-arrival rates and highly variant run-times of IIoT subsystems with low parallelism, thus improving energy efficiency.
\end{enumerate}
The above steps are repeated for developing an accurate global model for developing medical devices in healthcare industry. As individuals data is not shared on the cloud making FL as a trustable way to train the models.

\section{Challenges and Solutions}
\label{sec:Challenges}

The IIoT based systems are generally heterogeneous and the structure is also too complex in nature. Though these systems are advantageous, there are so many technical challenges in implementing and deploying these types of applications in real time. When these challenges are met by providing solutions, these systems will surely rule the world in near future. 

\subsection{\textcolor{black}{Effective Data Management}}
The IIoT systems are highly heterogeneous in nature and the data gathered using the data acquisition mechanisms or sensors and actuators are huge in volume and also at times unstructured. Also in IIoT systems, the data gathered are usually streams of data at high velocity. Currently these huge streams of data are stored in heterogeneous devices, gateways, edge servers, fog servers or cloud servers. The huge data gathered is raw and needs to be processed before taking any decisions in real-time \cite{diene2020data}. After processing, the processed data needs to be transmitted to the required destination in an efficient manner. The data once archived needs to be retrieved when requirement arises and hence the data has to be available at the right time for further analysis. While the data is archived, then arises the challenge for safe storage. Though IIoT is the buzz term in the recent days, most of the industries are still in a dilemma to implement these systems in real-time due to the above mentioned challenges.

\textbf{Solution:}
To cope up with these challenges, an efficient data management model \cite{zheng2020decentralized} has to be designed and utilized. The data model designed has to enable better storage mechanisms while archiving as well as sharing the sensitive data. It should also provide an efficient management scheme for handling the huge volume of data like processing, analyzing, retrieving the data at a higher speed,etc. 

\subsection{\textcolor{black}{Heterogeneity and Interoperability}}
IIoT \textcolor{black}{systems are typically comprised of} heterogeneous multiple vendor systems \cite{petroulakis2019semiotics} such as machines, robots, IoT devices, wearable sensors, actuators, networks which are either wired or wireless, 5G based cellular networks, broadband networks, storage servers which may be cloud servers or edge servers. Integrating these multiple technologies to build a whole system is a challenging issue.

\textbf{Solution:}
The researchers need to concentrate more on developing resource sharing techniques \cite{ramirez2019intelligent}, synchronizing techniques, data sharing techniques and similar factors among the heterogeneous environment to improve the interoperability \cite{platenius2020file} and collaboration.

\subsection{\textcolor{black}{Server side attack}} 

 Preserving privacy of data and model is major issue in decentralized framework. Security mechanism must be applied on Edge to end device communication and from edge to collaborative model. Even in worst case, the model updates can be tampered between end devices to collaborative engine. In FL, the update from end devices are sent to the cloud for training the global model. However there is a high chance of stealing the updates from cloud by the malicious user which results in increase of inconsistency and noise data in the model.
 
\textbf{Solution:} Fully Homomorphic Encryption (FHE) is one of the best to overcome privacy threat challenge in the cloud. All operation can be encrypted using FHE except the activation functions. Even for activation function, higher degree polynomials or modified Chebyshev polynomial can be applied and then can be encrypted using FHE. Recently Phong \emph{et al.} \cite{aono2017privacy} designed a model that balances the trade-off between data privacy and accuracy. It also ensures no information is leaked from the cloud server. The gradient updates are encrypted and is directly provided to neural networks where homomorphic function enables the computation among the other gradient updates.

\subsection{\textcolor{black}{Optimization} }
As the end devices have limited memory and computational requirements, DL models must be optimized so that they can be trained and deployed on IIoT or end devices efficiently. In terms of hardware optimization and model optimization would decreases computation time and energy consumption. To summarize. Optimizing the DL and ML models is an essential part to execute the models at the edge devices which remains a challenging task for the researchers even now. 
The end device have only limited power and resource to train the local model as they are shared by the other devices. Practically, it is difficult to provide a static connection for all end devices. So, we have to reduce the computational complexity and communication overhead in FL. Therefore, there should be a tradeoff between power consumption and resource allocation to train the model\cite{bonawitz2017practical}.

\textbf{Solution:} The main solution for the above challenge is to use MPC in FL by using secure aggregation protocol. So the best possible approach would be to use the FHE based MPC that can be implemented with limited rounds. Generally FHE provides the confidentiality and privacy of the updates but threshold FHE can reduce the communication overhead.

\subsection{\textcolor{black}{Inference Attack}}
 Inference attack is an attack where the hacker can infer sensitive information by the results of authorized query. Information can be inferred directly from the model or industry from the frequent queries. Generally the updates are sent to the server by the local node which indirectly has the features of the models. Some of the privacy information or identification of the user can also be inferred from the model updates. Inference attacks can be inferring class representatives, inferring membership, inferring properties and inferring training inputs and labels.There are two types of inference attack namely identification attack where the users  personal information is leaked and another type of attack is matching attack which compares the two model updates\cite{melis2018inference}.

\textbf{Solution:}
The best way is to utilize differential policy by adding noise to the information such that the unauthorized user cannot distinguish the original information from noise. Therefore there should be a balance between the privacy of data and accuracy of the model. This can be done by choosing the parameter to balance the above two parameters. Orekondy\cite{rahimian2020sampling} stated that domain specific data augmentation can provide effective results with minimal impact.   

\subsection{\textcolor{black}{Client side Attack}}
In general, privacy security mechanisms are computationally expensive techniques that results delay in response. Security issue on server side and client side remains a challenging task even the data communication is restricted. Some mechanisms namely  DP and homomorphic encryption technique controls the amount of data to be shared on the cloud.
Model poisoning is one of the dangerous attack which introduces backdoor functionality \cite{bagdasaryan2020backdoor} from the client side and poison the client model. Some of the popular backdoor attack model are label flipping attack\cite{tourani2017security} and Sybil attacks \cite{mishra2018analytical}. Based on the occurrence of their scenario, the complexity of the attack differs. One of the client device is compromised and the local model is trained using the backdoor data with their new techniques and submits the resulting model. Finally this poisoned client model is aggregated with other client model and replace the global model with the backdoor model. Another way of poisoning the model is the adversary can inject bad data into the models training data.

\textbf{Solution:} To prevent backdoor attack  combined anomaly detection algorithm, participant level  DP and Byzantine-tolerant gradient descent can be used. \cite{bonawitz2017practical} discussed about the secure aggregation as the updates from each end device are not visible to the aggregator. There is no one single solution to overcome the model poisoning. So it is difficult to design a best solution for various type poisoning techniques. Furthermore, various solution can be integrated into an automatic predictable model which is an open question to proceed research in future directions. Another recommended approach is participant level  DP. More specifically, FedAvg one of the  DP algorithm trains deep models on user partitioned data and another feature is providing required level of privacy for each participant.

\subsection{\textcolor{black}{Data Leakage}}
 The data generated by the smart devices are confidential, resulting in more security and privacy practices. Data leakage is one of the major barriers to data sharing in the distributed channel. Researchers are more focused on data leakage as sensitive data breaches can lead to increased risk factors like financial loss, online vandalism, future security costs. Providing secure, smart data transmission in a distributed environment is a major concern. 
 
 \textbf{Solution:}
Integrating FL and blockchain into IIoT can adequately improve data stabilization and data quality. Lu \emph{et al.} \cite{lu2019blockchain} proposed a blockchain-based FL model that enables data privacy in IIoT applications. The research aims to enhance trust management while transferring sensitive data from source to destination, to provide a secure blockchain-based data-sharing architecture in a distributed environment. The researchers pay special focus by establishing a blockchain environment to resist data leaks and inhibit full access control of it's owner. During this process, FL models are incorporated to conceptualize data models that often share data models in the communication network. However, smart mechanisms for enhancing the data utility and expanding the amount of resources for effective data sharing in a distributed setup should be incorporated.

\subsection{\textcolor{black}{Data Privacy}}
Data privacy is the primary concern in IIoT, as most of the data will be sensitive and confidential, so it is essential to keep these data secure. Data protection techniques are used to identify unauthorized users and malicious attacks to overcome these challenges. Data privacy is one of the key issue when FL is used to train static streams and data streams. In general, privacy can be classified as global and local privacy. The model updates generated in each iteration on all the devices are not available to untrusted third parties other than the centralized server, whereas in local privacy, the updates are private to the server. 

\textbf{Solution:}
In recent work \cite{arachchige2020trustworthy}, a PriModChain framework is proposed for the implementation of trust and data privacy in IIoT. The proposed model implemented  data privacy with a view to enhance the security and safety of the IIoT. The Smart Contract helps in providing transparency within the IIoT framework whenever a communication link is formulated between the distributed bodies and the central authorities. An interplanetary file system is used to achieve optimum latency, immutability, efficient data archiving and secure information sharing between the P2P network. Furthermore, the framework does not achieve optimum latency and thus curtails its use in the IIoT system.
Additionally, to improve resource optimization and data privacy for smart industries, the work in \cite{khan2020resource} unveils a dispersed federated approach for resource optimization and reliability. During this process, the proposed model aims to minimize FL costs, while optimization is achieved by a decomposing and relaxation-based algorithm. Later \cite{khan2020resource} indicated the convergence of sub-problems and remedied them using a stochastic optimization technique. 

\subsection{\textcolor{black}{Communication Overhead}}

Training in FL is a recursive process that involves intensive sharing of parameters among clients and servers. Consequently, to adopt FL to resource-constrained IIoT contexts, minimizing communication overhead is a key issue requiring in-depth investigation. However, some work has been undertaken to mitigate the use of resources in FL by lowering the load distribution factor between source and destination. 

\textbf{Solution:}
The work in \cite{wang2019adaptive} presented gradient-descent FL that involves local updates and global convergence measures. In order to accomplish the desired exchange between local update and global accumulation, a control algorithm is introduced to significantly reduce the loss function in the context of budgetary resource concerns. Meanwhile, researchers in \cite{nishio2019client} exemplify ineffective client training with fairly low resources. Besides, the lack of global accumulations could reduce model performance and convergence. The allocation of resources, data training and data authentication must be considered in order to improve the communication cost in FL.

\subsection{\textcolor{black}{Data Sharing in Horizontal Federated Learning}}
Industrial awareness is commonly observed using DL and data mining techniques related to IoT data. However some data is not easy to acquire one factory's information, as there are still few samples. If different alliance factories are able to gather their information together a lot more information could be extracted. In addition, security of information is a key issue for these factories. Traditional matrix-based techniques can ensure information security within a factory, but will not facilitate information sharing between factories, and therefore, due to lack of correlation, mining efficiency is weak. 

\textbf{Solution:} 
To achieve these objectives, a new Federated Tensor Mining (FTM) approach is introduced in \cite{kong2019federated} to federate multisource data for tensor-based mining while ensuring security. The key contribution of FTM is that every factory only needs to share its ciphertext data for security issue, and these ciphertexts are adequate for tensor-based knowledge mining due to its homomorphic attribution. Real-data-driven simulations demonstrate that FTM not only mines the same knowledge compared with the plaintext mining, but also is enabled to defend the attacks from distributed eavesdroppers and centralized hackers. 

\subsection{\textcolor{black}{Anomaly Detection}}
Effective anomaly detection techniques based on sensing big data become extremely important to ensure stability for industrial applications. Since the failure of edge devices (i.e. anomalies) has a serious impact on the production process in IIoT, the relevant and reliable detection of anomalies is becoming extremely important. In addition, the data collected by the edge device contains tremendous private information, which is challenging as user privacy has become a serious concern. 

\textbf{Solution:}
To address this issue, Liu \emph{et al.} \cite{liu2020deep} developed a communication-efficient FL-based deep anomaly detection model for IIoT data sensing. During this process, a FL model was first introduced to allow decentralised edge devices to train an anomaly detection model, which could enhance its classification accuracy. In the second step, an Attention Mechanism-based Convolutional Neural Network-Long Short Term Memory (AMCNN-LSTM) approach is formulated for accurate detection of anomalies. The AMCNN-LSTM method uses attention mechanism-based CNN units to identify relevant fine-grained features, to prevent memory loss and gradient dispersion risks. In addition, the proposed model retains the merits of the LSTM unit in anticipating time series information. In the third step, a gradient compression technique to improve communication process is proposed. 

In order to solve these problems, an innovative FL-based deep anomaly monitoring system is proposed for smart edge devices. Liu \emph{et al.} \cite{liucommunication} proposed a CNN-LSTM FL-based framework for detecting anomalies in IIoT applications.
To improve the data transmission efficacy of the presented framework, a gradient compression technique based on top-k selection has been used to minimise network overhead.

\subsection{\textcolor{black}{Data Integrity}}
Blockchain technology is used in IIoT-based FL to provide data integrity in order to obtain adequate customer information and training computing capabilities. Blockchain-based FL systems face some of the challenges in the design phase to address the barriers of data heterogeneity in IIoT during earlier fault detection. However, existing security mechanisms do not focus on designing the effective framework and blockchain's reliability issue.

\textbf{Solution:}
In \cite{zhang2020blockchain}, \textcolor{black}{a new FL-based blockchain architecture is proposed to detect failures in IIoT applications.} During this process, each client server continuously builds a Merkle tree where each leaf node signifies the client data and saves the root tree to the blockchain. An on-chain system is designed to measure each client's impact, based on the size of client data used in local model training.

\subsection{\textcolor{black}{Data and Model Attack}}

Even though data sharing is minimal in FL, still there can be several attacks in the framework at different levels.  In general attacks can happen at any stage starting from the basic node level to the server level. They are as follows:
\begin{enumerate}

\item Data Attack- Attack in local data at the end device  
\item Model Attack- Attack that happens in local trained model  and global model in centralised server

\end{enumerate}
 
\textbf{Data based attack}: Changing the data or labels used in the training data falls under this category of the attack. Data attack can be a clean label and dirty label attack. When the unauthorized person cannot change the label of data as this would lead to misclassification of data is known as clean label attack. In contrast in dirty label attack, the intruders attempt to include some number of samples which would mislead the ML process.

\textbf{Model based attack}: Changing the model updates communicated would cause the decrease in model efficiency. Hacking the model trained in local node results in leaking user identification data globally. Aggregated global model present in the centralized server is also prone to attack. Sometimes the model are targeted for misclassification by changing the training process. Even though DNN-based approaches have become more popular in IIoT applications, they are currently experiencing some disadvantages, such as massive numbers of adversarial attacks. How to build DNNs that are immune to different types of attacks are becoming a major bottleneck when deploying IIoT safety applications. 

\textbf{Solution:}
In order to provide solutions to these kinds of problems, Song \emph{et al.} \cite{song2020fda3} developed federated defensive system for cloud-based IIoT architectures. The revised FL model and the implemented adversarial learning loss function, the proposed strategy can metabolise DNNs to successfully handle existing adversarial threats while guaranteeing privacy protection among various IIoT devices. The obtained results were compared with two popular benchmarks, and the results reveal that the improved method can not only increase overall protection against various existing adversarial attacks, but can also accurately detect the improper behaviour of DNN induced by new attacks.

\section{Conclusion}
\label{sec:Conclusion}
 The new paradigm of integrating FL on IIoT data was surveyed and several approaches and techniques associated with it are introduced. In specific, we highlighted the characteristics and benefits of IIoT in terms of FL and distributive learning. The motivation behind the integration of IIoT with FL is also discussed. Subsequently, several privacy preserving ML/DL and blockchain models employed in FL and IIoT are presented. Then, survey on algorithms to handle heterogenous data is summarized. Contribution of researchers in the field of FL with data and resource management followed by several challenges and solutions are discussed. Reviewing automobile and healthcare applications on IIoT with FL would provide the researchers to understand the principal components of IIoT smart devices. We also identified various challenges, solutions and future research direction in path of FL for IIoT. In summary, the advantages of extending edge intelligence to intelligent edge is to reduce the communication cost, computing power and energy consumption. Also, this paper would enhance the readers to consolidate and combine the information with respect to IIoT, FL and privacy preserving mechanism, data management and storage mechanisms under a common new paradigm of Industry 4.0. 
 
 In future, researchers should focus on developing algorithms and methodologies to impose privacy constraints across edge device and data level. The privacy preserving model for FL should be designed to learn multiple geographically distributed training data sets without exploiting the sensitive information of the user. The existing architecture violates the usage of short packets for low latency communication. Therefore new generation of wireless systems such as 6G cellular networks can be integrated with FL enabled IIoT networks to provide more importance for privacy preservation.


\end{document}